\documentstyle[preprint,aps,eqsecnum]{revtex}
\tightenlines

\def\bomeg{\pmb{$\omega$}}
\def\bbomeg{\pmb{$\Omega$}}

\def\bPhi{\pmb{$\Phi$}}

\def\bomeg{\pmb{$\omega$}}
\def\bbomeg{\pmb{$\Omega$}}

\def\bPhi{\pmb{$\Phi$}}

\def\beq{\begin{equation}}
\def\eeq{\end{equation}}
\def\bea{\begin{eqnarray}}
\def\eea{\end{eqnarray}}
\begin{document}
\title{
Semiclassical theory of surface plasmons in spheroidal clusters}
\author{A. Dellafiore$^{a}$, F. Matera$^{b,a}$ and F. A. Brieva$^{c}$}

\address{$^{a})$ Istituto Nazionale di Fisica Nucleare,
                    Sezione di Firenze,}
\address{$^{b})$ Dipartimento di Fisica,
              Universit\`a degli Studi di Firenze,\\
             L.go E. Fermi 2, I-50125, Firenze, Italy}
\address{$^{c})$ Departamento de F{\'\i}sica,
         Facultad de Ciencias F{\'\i}sicas y  Matem\'aticas,\\
        Universidad de Chile, Casilla 487--3, Santiago, Chile}
\maketitle

\begin{abstract}
A microscopic theory of linear response based on the Vlasov equation
is extended to systems having spheroidal equilibrium shape.
The solution of the linearized Vlasov equation, which gives a semiclassical
version of the random phase approximation, is studied for electrons moving
in a deformed equilibrium mean field. The deformed field has been
approximated by a cavity of spheroidal shape,
both prolate and oblate. Contrary to spherical systems,
there is now a coupling among excitations of different
multipolarity induced by the interaction among constituents. 
Explicit calculations are performed for the dipole response of 
deformed clusters of different size.
In all cases studied here the photoabsorption strength for prolate clusters
always displays a typical double-peaked structure. For oblate clusters
we find that the high--frequency component of the
plasmon doublet can get fragmented 
in the medium size region ($N \sim 250$).
This fragmentation is related to
the presence of two kinds of three-dimensional electron
orbits in oblate cavities.
The possible scaling of our semiclassical equations with the
valence electron number and density is investigated.
\end{abstract}

\section{Introduction}

	Complex many-fermion systems like atomic clusters and nuclei
exhibit both spherical and deformed equilibrium shapes. Deformation of the
ground state gives rise to observable effects in the excitation spectrum
of these systems,  with the splitting of the giant dipole resonance in 
deformed nuclei most likely being the best well known feature\cite{danos}.
Based on the close analogy that exists between the nuclear giant dipole 
resonance and the cluster surface plasmon, a similar splitting is expected
and has indeed been observed in atomic
clusters (see Refs. \cite{1} and \cite{brack} for reviews of experimental
and theoretical work).  A spheroidal deformation (prolate and oblate)
is often sufficient to explain data in both nuclei and atomic clusters
although recent jellium model calculations \cite{2}  also suggest more 
complicated shapes for the latter.

Clemenger \cite{3} has used a deformed oscillator model,
which is inspired by the Nilsson model of nuclear physics \cite{4},
to describe deformed clusters.  However Strutinsky $et~al.$ \cite{5} have
pointed out that the oscillator potential is somewhat special and argued 
that a spheroidal cavity would give a more realistic description of the 
equilibrium mean field (in nuclei, but the same is true for large
atomic clusters). 
Actually the spheroidal cavity is still a rather special choice since its
sharp surface is an undesirable feature. A mean field with a diffuse 
surface  would be more realistic. Studies by Arvieu $et~al.$ \cite{6} 
of the classical motion of a particle in a deformed Saxon--Woods like 
potential indicate that part of the classical phase space becomes chaotic.
In a spheroidal cavity instead the single-particle motion is integrable and
this fact makes calculations practicable.

In this paper we study  surface plasmons in deformed clusters by 
assuming a spheroidal cavity model for the equilibrium mean field.
Thus we limit our analysis to the simplest deviation from spherical
shape and consider only spheroidal geometry. However our model
is not limited to small deformations and, in principle,
we can study systems ranging from spherical to almost cigar or 
disk-like shapes. Our aim is to extend the semiclassical
theory of linear response based on the Vlasov equation of Refs.\cite{7,8}
to deformed systems. This theory can be viewed as a semiclassical 
version of the random phase approximaton (RPA). Although a fully 
quantum treatment of the problem is certainly more rigorous,
the Vlasov equation has the advantage that the numerical effort 
required is greatly reduced. Indeed, a fully quantum RPA 
calculation for finite systems must rely heavily on numerical
computation. This fact has limited  explicit calculations  mostly to
either infinite homogeneous  or  spherical systems, where the
symmetries of the mean field Hamiltonian allow for simplifications
of the numerical problem. Pioneering work on the quantum response of
spheroidal clusters has been made by  Ekardt and Penzar \cite{9}
(for prolate clusters only).

The present work is organized as follows. In Sect. II we extend the 
formalism of Refs.~\cite{7,8} to spheroidal systems. In order to present
clearly the main points of the theory we have relegated the discussion
of many detailed expressions to the Appendix. In Sect. III the model is 
applied to study the evolution of the peak profile of surface plasmons 
with deformation. Both the "single-particle" and collective responses 
are studied for prolate and oblate geometries. Finally, Sect. IV 
contains a brief summary and the conclusions.
\section{Formalism}

In spite of its well known difficulties in reproducing the
observed width of collective resonances, the RPA is still the basic
microscopic theory of small--amplitude vibrations for many fermions 
systems. The RPA theory can be derived in the Green function 
formalism \cite{10}. The RPA Green function $G$ obeys an integral 
equation that can be written formally as
\beq
\label{1.1}
G(\omega) = G^{0}(\omega)+G^{0}(\omega)VG(\omega)\:,
\eeq
where the unperturbed particle-hole Green function $G^0$ describes the
single-particle motion of the constituents in the equilibrium mean field, 
and $V$ is the effective two-body residual interaction.
The response of the many-body system to a weak periodic external field of 
frequency $\omega$ is proportional to the imaginary part of $G(\omega)$.
In practice, the calculation of $G$ from Eq.~(\ref{1.1}) means
facing two main problems: first to evaluate $G^0$ for the system under 
study  and then solving the corresponding integral equation to determine $G$. 
The latter becomes quite a challenge when the system is non--spherical 
since $G$ will be determined by a system of coupled integral equations, as 
we shall discuss later on.

The problem of calculating $G^{0}$ for a given mean field is simpler in 
the semiclassical theory of linear response developed in 
Refs.~\cite{7,8}. This theory has been applied to
the study of giant resonances in spherical nuclei \cite{12} and of
surface plasmon excitations in spherical microclusters \cite{13}.
The excitation spectra given by this semiclassical theory are very similar
to those yielded by the fully quantum RPA theory even when the many-body
system under investigation is not particularly large.
Moreover for large deformations it is expected (Ref.~\cite{4}, p.591) 
that shell effects should be smaller than for spherical systems, thus 
favouring a semiclassical approach.

The semiclassical theory of linear response based on the Vlasov equation 
of Refs.~\cite{7,8} leads to an expression for the propagator
with the same structure as the quantum RPA given by Eq.~(\ref{1.1}). 
However the RPA equation is actually more complicated when exchange (Fock) 
terms are properly taken into account \cite{10}. Nevertheless, these terms 
are often treated in a local approximation leading to the same Hartee--like 
structure as the Vlasov equation. In the classical limit, $G$ and $G^0$ 
will be denoted by $D$ and $D^0$ respectively in order to distinguish them 
from the corresponding quantum quantities.  

Although the theory of Refs.~\cite{7,8} is generally valid for all integrable 
mean field Hamiltonians, practical applications have been limited
to spherical systems. In order to extend its range of application 
to deformed systems we consider the classical limit of Eq.~(\ref{1.1})
in momentum space,
\beq
\label{2.1}
D({\bf q}^{\prime},{\bf q},\omega)=D^{0}({\bf q}^{\prime},{\bf q},\omega)+
\frac{1}{(2\pi)^{3}} \int d{\bf k}\,D^{0}({\bf q}^{\prime},{\bf k},\omega)\,
V(k)\,D({\bf k},{\bf q},\omega)\:.
\eeq
Following \cite{7,8}, and using units $\hbar=c=1$, the propagator $D^{0}$ 
can be written in terms of action--angle variables $\{{\bf I},\bPhi\}$ as 
\beq
\label{2.4}
D{^0}({\bf q}^\prime,{\bf q},\omega)=(2\pi)^3
\sum_{\bf n}\int d{\bf I}\,F'(h_{0}({\bf I}))\,
\frac{{\bf n}\cdot {\bomeg}({\bf I})}
{{\bf n}\cdot {\bomeg}({\bf I})-(\omega +
i\varepsilon)}\,Q^{*}_{\bf n}({\bf q}',{\bf I})~Q_{\bf n}({\bf q,I})\:,
\eeq 
with $\varepsilon \rightarrow 0$, and the Fourier coefficients
\beq
\label{2.5}
Q_{\bf n}({\bf q,I})=\frac{1}{(2\pi)^3}\int d\bPhi\,e^{-i{\bf n}\cdot{\bPhi}}\,
e^{i{\bf q}\cdot{\bf r}}\:,
\eeq
taking the place of the quantum matrix elements (see also \cite{14}).
In Eq.~(\ref{2.4}), {\bf n} is a three-dimensional vector with integer
components and the sum extends to all possible values of {\bf n}. 
The components of the vector $\bomeg$ are the fundamental frequencies 
of the multiply--periodic particle motion in the equilibrium mean field,
\beq
\label{omegas}
{\bomeg}({\bf I})= \nabla_{I} h_0({\bf I})\:,
\eeq
with $h_{0}=E$ the equilibrium Hamiltonian. The function $F'(E)$ 
in Eq.~(\ref{2.4}) is the derivative of
\beq
\label{fzero}
F(E)={{2}\over{(2\pi\hbar)^3}}\, \theta(E_F -E)\:,
\eeq
which describes the equilibrium distribution of electrons at zero 
temperature ($E_F$ is the Fermi energy). Thus,
\beq
\label{2.8}
F'(E)=-\,\frac{2}{(2\pi\hbar)^3}\, \delta(E_F -E)\:,
\eeq
expression that reduces by one the number of integrals in Eq.~(\ref{2.4}).
We also note that
the  propagator in Eq.~(\ref{2.4}) is perfectly well behaved when
${\bf n}\cdot \bomeg \to 0$ due to the presence
of the  ${\bf n}\cdot \bomeg$ factor in the numerator. Then $D^0$
is not affected by the problem of small divisors (Ref.~\cite{15},
p.523).

For a realistic mean field the frequencies $\bomeg$ do depend upon 
the value of the action variables ${\bf I}$. This dependence reflects the
non-linearity of the mean field. A cavity is an example where
this non-linearity effect is present.
Instead, for the oscillator potential model where 
the equilibrium Hamiltonian is given by
\beq
\left(\,h_0\,\right)^{osc} =\bomeg \cdot {\bf I} \:,
\eeq
the frequencies $\bomeg$ do not depend on ${\bf I}$. This difference 
is a basic distinction between the oscillator and  more realistic 
models  and it becomes  the main reason for our choice to study the 
cavity model.

The most general partial--wave expansion of the propagator
$D{^0}({\bf q}^\prime,{\bf q},\omega)$ is
\beq
D^0 ({\bf q}^{\prime},{\bf q},\omega)=\sum_{LM}~\sum_{L'M'}
D^{0}_{LM,L'M'}(q^\prime,q,\omega)
Y_{L^\prime M^\prime}({\hat{\bf  q}}^{\prime})Y_{L M}^*({\hat{\bf q}})\:.
\eeq
The symmetry properties of the system allow for a simplification of
this expression. Indeed, the axial symmetry of spheroidal 
(and also spherical) systems implies
\beq
D^{0}_{LM,L'M'}=\delta_{_{M,M'}} D^0 _{L'LM}(q',q,\omega)\:,
\eeq
and therefore
\beq
\label{exp}
D^0 ({\bf q}^{\prime},{\bf q},\omega)=\sum_{L=0}^{\infty} \sum_{M=-L}
^{L} \sum_{L^\prime=|M|}^{\infty}D^0 _{L^\prime L M}(q^\prime,q,\omega)
Y_{L^\prime M}({\hat{\bf  q}}^{\prime})Y_{L M}^*({\hat{\bf q}})\:.
\eeq

A similar expansion for the propagator $D({\bf q}^{\prime},{\bf q},\omega)$ 
is obtained from Eq.~(\ref{2.1}). Thus, for spheroidal systems, 
the three--dimensional integral equation (Eq.~(\ref{2.1})) reduces to the 
following system of coupled one--dimensional integral equations
for the coefficients $D_{L^\prime L M}(q^\prime,q,\omega)$,
\begin{eqnarray}
\label{2.13}
D_{L^\prime L M}(q^\prime,q,\omega)&=&
D^{0}_{L^\prime L M}(q^\prime,q,\omega) \nonumber \\
& &+\,\frac{1}{(2\pi)^3} 
\sum_{\ell=|M|}^{\infty} \int_{0}^{\infty}dk\,k^2
D^{0}_{L^\prime \ell M}(q^\prime,k,\omega) V(k)
D_{\ell L M}(k,q,\omega)\,. \nonumber \\
\end{eqnarray}
This equation can be further simplified for
spherical symmetry. In this case,
\beq
D^0 _{L^\prime L M}(q^\prime,q,\omega)=\delta_{L,L'}D^0 _L (q',q,\omega)
\qquad  ({\rm any}~M) \:,
\eeq
and Eq.~(\ref{2.13}) reduces to a single uncoupled one--dimensional
integral equation for each partial--wave component,
\beq
\label{2.9}
D_{L }(q^\prime,q,\omega)=D^{0}_{L }(q^\prime,q,\omega)
+{1\over (2\pi)^3}\int_{0}^{\infty}dk\,k^2
D^{0}_{L}(q^\prime,k,\omega) V(k)
D_{L }(k,q,\omega)\:.
\eeq
The solution of this equation for surface plasmons in spherical clusters has
been studied in Ref.~\cite{13}, where the coefficients $D^0 _L (q',q,\omega)$
have been explicitly derived (see also Ref.~\cite{8}).

In order to solve the system of coupled integral equations expressed
by Eq.~(\ref{2.13}), we must derive an explicit expression for the 
coefficients $D^0 _{L^\prime L M}(q^\prime,q,\omega)$. This can be done 
in a way similar to that followed in Ref.~\cite{7}. We refer here to the 
Appendix
for details on how to extend that approach to both prolate
and oblate geometries. From Eqs.~(\ref{2.4}) and (\ref{exp}) we obtain
\begin{eqnarray}
\label{dzero}
D^0_{L^{\prime}LM}(q^\prime,q,\omega)&=&-\,\frac{2}{(2\pi\hbar)^3}
\sum_{n_u,n_v} (2\pi)^3
\int d\lambda_z\int d\epsilon
\left|\,\frac{\partial(I_v,I_u)}{\partial(E,\epsilon)}\,\right| 
\nonumber \\
& &\,\times \frac{n_u \omega_u +n_v \omega_v +M\omega_{\varphi}}
{n_u \omega_u +n_v \omega_v +M\omega_{\varphi}-(\omega+i\varepsilon)}\nonumber 
\\
& &\,\times Q^{(L'M)*}_{n_u,n_v,M}(E,\epsilon,\lambda_z;q^\prime)\,
Q^{(LM)}_{n_u,n_v,M}(E,\epsilon,\lambda_z;q)\:, 
\end{eqnarray}
where we have used the constants of motion $\{E,\epsilon,\lambda_z\}$ instead
of  the action variables ${\bf I}$ to evaluate the corresponding integrals. 
All vector quantities in Eqs.(\ref{2.4}) and (\ref{dzero}) are
expressed through their  $(u,v,\varphi)$ components
as discussed in the Appendix.
Furthermore, the integrand is evaluated at the Fermi
energy $E_F$ as a consequence of Eq.~(\ref{2.8}). All other integrals are
extended to the classically available phase space, the integration limits
are specified in the Appendix.

The Fourier coefficients $Q^{(LM)}_{n_u,n_v,M}(E,\epsilon,\lambda_z;q)$ 
are related to those of Eq.~(\ref{2.5}) by
\beq
Q_{\bf n}({\bf q},{\bf I})=\sum_{LM}Q_{\bf n}^{(LM)}Y_{LM}^*(\hat{\bf q}),
\eeq
and
\begin{eqnarray}
\label{fou}
Q^{(LM)}_{n_u,n_v,n_\varphi}&=&{1\over{\pi}}\,
\delta_{M,n_\varphi}\oint du\oint dv \left|\frac{\partial (\Phi_u,\Phi_v)}
{\partial (u,v)}\right| e^{-i(n_u\Phi_u+n_v\Phi_v+n_\varphi
\tilde\Phi_\varphi)} \nonumber \\
& &\times\,i^L\,Y_{LM}(\theta (u,v),0)\,j_L(qr(u,v))\:.
\end{eqnarray}
The symbol $\oint$ means integration over a whole period of
classical motion in the respective variable, the angle variable  $\Phi_\varphi$
takes the form $\Phi_\varphi = 
\tilde\Phi_\varphi(u,v) + \varphi$  and $j_L$ is
the spherical Bessel function of order $L$.

The comparison of Eqs.(\ref{2.13}) and (\ref{2.9}) explicitly shows that,
contrary to the spherical case,  there
is a coupling between excitations of different multipolarity 
in deformed systems.
It can be shown (see Appendix) that $D^0_{L'LM}=0$ unless 
$(-)^{L'}=(-)^L$, so that only multipoles with the same parity are 
mixed in Eq.~(\ref{2.13}). Since this is a consequence of
the reflection symmetry of the spheroid ($z \to -z$), it holds
both for prolate and oblate shapes. Such a simplification
would not necessarily occur for "pear-shaped" systems.

The physical observables we consider in this work are related to the
"forward" propagator $D({\bf q},{\bf q},\omega)$ regardless of the need
for the full off--diagonal propagator $D({\bf q}^{\prime},{\bf q},\omega)$
to solve Eq.~(\ref{2.1}).  Moreover, for randomly oriented clusters, we must 
average over the solid angle \cite{danos}. Thus we define the 
angle-averaged propagator
\beq
<D(q,\omega)>\equiv {{1}\over{4\pi}}\int d{\hat{\bf q}}\,
D({\bf q},{\bf q},\omega)\:.
\eeq
The expansion  of $D({\bf q}^{\prime},{\bf q},\omega)$ analogous
to Eq.~(\ref{exp}) implies
\beq
<D(q,\omega)>=\frac{1}{4 \pi}\,\sum_{LM}D_{LLM}(q,q,\omega)\:.
\eeq

A quantity often used in cluster physics is the dynamic polarizability
$\alpha(\omega)$. It is related to the longwavelength limit of the $L=1$
component of $<D(q,\omega)>$. For deformed clusters the polarizability is a
tensor since the induced dipole moment depends on the orientation of the
cluster relative to the external field. We introduce the $M$--component
polarizability as
\beq
\label{alpham}
\alpha_M (\omega)=-\,{\textstyle\frac{1}{3}}\,e^2 \lim_{q\to 0}\left\{
\frac{1}{4 \pi}\,{{D_{11M}(q,q,\omega)}\over{[\,j_1 (q)\,]^2}}\right\}\:.
\eeq
In the spherical limit $\alpha_M$ is $M$--independent.
For spheroidal systems $\alpha_{-M}=\alpha_M$ (this is a property 
of the coefficients $D_{L'LM}$).

We introduce also the "partial" photoabsorption cross sections
\beq
\label{sigmam}
\sigma_M (\omega)={{4\pi}\over c}\,\omega\,{\rm Im}\left[\,\alpha_M (\omega)\,
\right]\:,
\eeq
and the total photoabsorption cross section is \cite{9}
\begin{equation}
\label{2.22}
\sigma(\omega) ={\textstyle\frac{1}{3}}\,\sum_{M=-1}^1 \sigma_M (\omega)
={\textstyle\frac{1}{3}}\,
\left[\,\sigma_0 (\omega) +2\,\sigma_1 (\omega)\,\right]\:.
\end{equation}

\section{Results}

In this section we apply the formalism developed in Sect. II and in
the Appendix to the study of prolate and oblate clusters 
with varying degree of deformation.
Also we explore the possibility of scaling with the number
of valence electrons and cluster density.
The emphasis is on understanding how the collective response is altered  
when the clusters under investigation do not have a spherical shape. 
We shall not attempt to compare our results with experimental
data at this level of development. For this reason we do not consider
such refinements as the electron "spill out" which is well known to 
give a red shift of the plasmon peak in spherical clusters. We expect 
to find a similar effect in the deformed case.
No particular effort is made to reproduce the observed width
of the plasmon resonances which is considerably larger than that obtained in
the present model.

The picture we have in mind here corresponds to taking a spherical cluster
and then deforming it to either a prolate or oblate shape while keeping 
its volume and density constant. We consider initially a relatively 
large sodium cluster of spherical shape, with $N=254$ valence electrons, 
described approximately by a square-well mean-field potential of radius 
$R=r_{s}\,N^{1\over 3 }$ and $r_s $ is the Wigner-Seitz parameter in units 
of the Bohr radius. We will take $r_s =4.0$. In this case the 
surface plasmon resonance consists of a single peak situated near the 
Mie frequency,
\begin{equation}
\label{omegamie}
\omega_{Mie} = \frac{\omega_p}{\sqrt{3}}\:,
\end{equation}
with $\omega_p $ the bulk plasmon frequency. Then, this sodium cluster
is deformed to a spheroidal (prolate or oblate) shape characterized by
a deformation parameter $\eta$,
\begin{equation}
\eta = \frac{R_{>}}{R_{<}}\:,
\end{equation}
where $R_{>}$ $(R_{<})$ are the larger (smaller) diameters of the spheroids.
We study the changes that are expected in the distribution of the dipole 
strength and  report results for $1 \le \eta \le 2$. The spherical limit is
obtained by taking $\eta = 1.001$.

To describe the spheroidal systems we introduce spheroidal coordinates
$\{u,v,\varphi\}$ as described in the Appendix for the prolate and oblate
geometries. Assuming that electrons move in a static cavity of spheroidal
shape, the equilibrium Hamiltonian $h_0$ will be chosen to be
\begin{equation}
h_0(u,v,\varphi,p_u,p_v,p_{\varphi}) = K(u,v,\varphi,p_u,p_v,p_{\varphi}) +
{\mathcal V}(v)\:,
\end{equation}
where $\{p_u,p_v,p_{\varphi}\}$ are the corresponding conjugate momenta, $K$ is
the kinetic energy and ${\mathcal V}$ the potential energy for particles in
a cavity. All the calculations have been performed within this framework.
The central objective has been to calculate the collective $D_{1 1 M}$ 
function from Eq.~(\ref{2.13}). The number of coupled integral equations
required is 3 for the cases of larger deformation ($\eta = 2$)  
considered in this work (that is $\ell =1,3,5$ in Eq.~(\ref{2.13})). 

One technical point deserving special 
attention is the handling of the singularities present in the $D^0$ 
propagator in Eq.~(\ref{dzero}). We have taken a finite value 
$\varepsilon = 0.002\, \omega_{Mie}$ when performing the actual
calculations. Our choice  modifies slightly ($\sim 5\%$) the width 
of either the single--particle or the collective plasmon peak. Furthermore,
we have summed up all allowed frequency modes for 
$-\,2 \le \{n_u,n_v\} \le 2$ when calculating $D^0$. The dynamic
polarizability and the total photoabsorption cross section of prolate clusters
are mainly dominated by
the $n_u=1$, $n_v=0,\pm 1$ modes for $M=0$ and $n_u=0$, $n_v=0,\pm 1$ for $M=1$.
However, the oblate cavity presents some peculiar features requiring
a finer analysis.

From a theoretical point of view it is interesting to study also the
single-particle response (proportional to the imaginary part of the zero-order
propagator $D^0 $) since its features are more directly related to the
shape of the equilibrium mean field. This can be of help in understanding
how this shape affects the collective response.

The effective two-body residual interaction $V(k)$ in Eq.(\ref{2.1})
determines primarily the position of the collective plasmon peak and, 
to a lesser extent, its shape. The present semiclassical
approach is essentially a Hartree approximation which does not
include exchange contributions. However, exchange and correlation terms can
be taken into account in a local approximation by introducing a (static)
local-field correction $G(k)$ \cite{16}. Thus the momentum-space interaction
we use is
\beq
\label{force}
V(k)=4\pi\,{{e^2}\over{k^2}}\,\left[\,1-G(k)\,\right]\:,
\eeq
with 
\beq
G(k)=A\,\left[\,1-e^{-\,B\,(k/p_F)^2}\,\right]\:,
\eeq
where $A = 0.9959$ and $B = 0.2612$ for sodium  
(\cite{16}, p.446)
and $p_F$ is the Fermi momentum. In our calculations we have verified that
it is possible to take an upper integration limit $k_{max}=4p_F$ when
solving Eq.(\ref{2.13}) without a noticeable change in the relevant results.

\subsection{Prolate cavity}

The natural frequencies $\{\omega_u,\omega_v,\omega_{\varphi}\}$ of the 
unperturbed
trajectories of particles in the prolate cavity determine the gross behaviour
of the semiclassical propagator $D^0$ (Eq.~(\ref{dzero})) as a function of 
$\omega$.
Since these frequencies depend on the integration variables $\epsilon$ and
$\lambda_z $ it is 
interesting to evaluate the possible values that they take and their
occurrence as a function
of the cluster deformation $\eta$. In Fig.~\ref{NF1} we show histograms for the
probability density of the natural frequencies expressed in terms of the Mie
frequency $\omega_{Mie}$ and for deformations $\eta = 1.0$, $1.5$ and $2$.
This probability density refers to the occurrence  of the
frequencies as functions of the (discretized) integration parameters $\epsilon$
and $\lambda_z$ in our numerical calculations. It
was evaluated with bins of width $0.002\:\omega_{Mie}$.
The frequencies $\omega_u$ and $\omega_{\varphi}$ fall in a narrow
range of the order of $0.05\:\omega_{Mie}$, with the former moving to the
left and the latter to the right of the spherical limit as deformation
increases. For $\omega_{\varphi}$ we have not plotted the 
symmetric negative components
in Fig.~\ref{NF1}. The frequency
$\omega_v$ shows a different behaviour and it spans, in principle, the
$[\sim 0.5\:\omega_{Mie},\infty]$ range. Overall, Fig.~\ref{NF1} gives a clear 
idea of the dominant frequency poles contributing to the $D^0$ propagator.

The $M=0,1$ components of the photoabsorption cross section 
(Eq.~(\ref{sigmam})) per valence electron are shown in Fig.~\ref{NF2}
for prolate clusters with deformation ranging from $\eta = 1.0$ 
to $2.$, step $0.25$. We display the results in a 3--dimensional
plot to better assess their relative behaviour.
The areas shaded black 
show the cross section obtained from
the zero--order propagator $D^0$. 
As a reference, we can observe on the left--hand side 
of Fig.~\ref{NF2} the same peak 
corresponding to spherical geometry ($\eta=1.0$) 
in both $M=0$ and $1$ components 
and centered around $\omega \simeq 0.16\:\omega_{Mie}$. Then, the
$M=0$ component shows a photoabsorption peak shifted to lower 
frequencies as deformation increases. 
This is a consequence of 
$\omega_u$ (see Fig.~\ref{NF1}) determining the dominant pole.
The corresponding 
strength of the peak increases for $\eta = 1$ to $\sim 1.5$ and then
slowly starts to decrease. 
The $M=1$ peak is shifted to higher frequencies due to the
dominance of  $\omega_{\varphi}$ in this mode and
its strength decreases with increasing deformation.

The right--hand side of  Fig.~\ref{NF2} shows the partial collective cross sections
(shaded gray) evaluated  from the full 
propagator $D$ in Eq.~(\ref{alpham}).
By including the fluctuation of the mean field  we obtain two main
effects: a huge shift in the position of the $M$ component of the
photoabsorption peak to frequencies around the Mie frequency and 
a noticeable change both in the width and the strength of the peak, 
mainly for $M=1$. We have numerically checked that the energy--weighted
sum rule (area under the curves) is unchanged within $2\%$.
The low--energy collective plasmon ($M=0$ component) has an intrinsic
structure which is slightly more complex than the corresponding
high--energy plasmon ($M=1$ component) and showing some degree of
fragmentation noticeable at $\eta \sim 1.5$.
The dominant collective peaks in  Fig.~\ref{NF2}
are slightly blue shifted with respect to the positions
predicted by the  Mie theory both for the $M=0$ and $1$ components.
This completely classical theory predicts that
the $M=0$ and $M=1$ peaks should be at the frequencies $\omega_{i}$
corresponding to oscillations along the $i$-axis ($z$ for $M=0$,
$x$ and $y$ for $M=1$). These frequencies are given by
\beq
\label{3.3}
\omega_i =\sqrt{n_i}\,\omega_p\:,
\eeq
with $n_i$ the appropriate
depolarizing factor. For spherical symmetry $n_x =n_y =n_z =\frac{1}{3}$, giving
Eq.~(\ref{omegamie}), while
for a prolate spheroid \cite{17}
\begin{eqnarray}
\label{3.4}
n_x &=& n_y = \textstyle{\frac{1}{2}}\,(1-n_z)\:, \\
n_z &=& {{1-e^2 }\over{2e^3 }}\left[\,\log \left(\, \frac{1+e}
{1-e}\,\right)-2e\,\right]\:.
\end{eqnarray}
The eccentricity $e$ is related to our deformation parameter $\eta$ by
\begin{equation}
e=\sqrt{\,1-\textstyle{\frac{1}{\eta^2}}\,}\:.
\end{equation}
A simple calculation gives $\omega_{x,y} = 1.073\:\omega_{Mie}$,
$~\omega_z = 0.836\:\omega_{Mie}$ for $\eta = 1.5$ and 
$\omega_{x,y} = 1.113\:\omega_{Mie}$,
$~\omega_z = 0.722\:\omega_{Mie}$ for $\eta = 2$. From Fig.~\ref{NF2}
we observe that our collective plasmon peaks 
are blue shifted by $\sim 7 \%$ with respect to the position 
expected from the classical values.

\subsection{Oblate cavity}

We turn now to present our results for oblate sodium clusters.
One of the most interesting aspects of the present cavity model is the
existence of two kinds of three-dimensional orbits, $W$ and $B$ orbits,
for oblate geometry (see Appendix). This feature should be shared,
at least qualitatively, by more realistic deformed mean fields.
In order to estimate the relative importance of the two kinds
of orbits,
we have plotted in Fig.~\ref{NF3} the fraction of valence electrons 
following each of the orbits as a function of the
cluster deformation $\eta$. In the spherical limit, $\eta \to 1$, the
cavity only allows $W$--type three--dimensional orbits since the
$B$ orbits become oscillations along a diameter.
As deformation increases the fraction of electrons in $B$ orbits raises
quickly reaching $50\%$ for $\eta \simeq 1.7$.
The presence of  $W$ and $B$ orbits 
in the oblate cavity
suggests that we may expect a richer behaviour of the cluster response.

The $W$ and $B$  orbits are characterized by their respective
fundamental frequencies (Eq.~(\ref{omegas})) which in turn
determine the $D^0$ propagator.
We shall refer to them as the $\{\omega^{W}\}$ and $\{\omega^{B}\}$
frequencies. In order to have an estimate of the possible values 
they take and of
their occurrence for different deformations of the oblate cluster we present
histograms for the probability density of the natural frequencies 
$\{\omega_u^{W},\omega_v^{W},\omega_{\varphi}^{W}\}$ in Fig.~\ref{NF4}
and of $\{\omega_u^{B},\omega_v^{B},\omega_{\varphi}^{B}\}$ in Fig.~\ref{NF5}.
The $\{\omega^{W}\}$ frequencies in Fig.~\ref{NF4}
may be compared to the natural frequencies
obtained for the prolate cavity shown in Fig.~\ref{NF1}. 
Clearly
the results for $\eta = 1$ in both cases are identical.
A great similarity is also observed for
$\omega_v$ as the clusters get deformed. However, the increasing 
occurrence of $\omega_v$ in Fig.~\ref{NF1} at the lower
limit  of the frequency range is
not present in $\omega_v^{W}$ (Fig.~\ref{NF4}).
The allowed values for $\omega_u^{W}$ change 
slightly with deformation and are not sharply defined. The
behaviour of $\omega_u$ in Fig.~\ref{NF1} is rather different,
with the frequency defined in a narrower range and its allowed values
decreasing with increasing deformation of the prolate cavity. The
$\omega_{\varphi}^{W}$ frequency in Fig.~\ref{NF4} shows a similar structure to
its counterpart in the prolate case but its magnitude decreases with 
deformation rather than increasing (Fig.~\ref{NF1}).

The $\{\omega^{B}\}$ frequencies are characterized by a different
probability density , as shown in
Fig.~\ref{NF5}. In the spherical limit ($\eta = 1$) the $B$ orbits have
perfectly defined frequencies and $\omega_u^{B} = \pm\,\omega_{\varphi}^B =
{\textstyle \frac{1}{2}}\,\omega_v^{B}$. As deformation increases
$\omega_u^{B}$ becomes much less defined and, on average, takes smaller
values than in the spherical case. The frequency  $\omega_{\varphi}^{B}$ instead
is still fairly well defined and is also decreasing with increasing 
deformation. Furthermore, $\omega_{\varphi}^{B}$ shows great similarity with
$\omega_{\varphi}^{W}$
in Fig.~\ref{NF4}. The frequency $\omega_v^{B}$ shows a definitely different behaviour
than $\omega_v^{W}$. It spans a finite range of values depending
strongly on the cluster deformation.
From the results shown in Figs.~\ref{NF4} and \ref{NF5}
we may expect a more complex behaviour of the $D^0$ propagator
than that observed in the prolate cavity.

The $M=0,1$ components of the photoabsorption cross section 
(Eq.~(\ref{sigmam})) per valence electron are shown in Fig.~\ref{NF6}
for oblate clusters with deformation ranging from
$\eta=1.0\, (0.25)\, 2.$.
As in Fig.~\ref{NF2} we have plotted the cross sections 
$\sigma_M^{(0)}$ (shaded black) obtained from the propagator $D^0$ calculated
in the static mean field and $\sigma_M$ (shaded gray) 
calculated from the correlated propagator $D$.
For $M=0$ we observe $\sigma_0^{(0)}$ presenting a novel structure.
Indeed, the  dipole peak is splitted into several peaks as deformation
of the oblate cluster increases. A simple estimate based on the probability
densities for $\omega^W$ and $\omega^B$ indicates that both $W$ (dominant)
and $B$ orbits contribute to the left most peak in the mode
$\{n_u=1,n_v=0\}$.  At higher
frequencies the other two peaks come from $B$ orbits in modes
$\{n_u=-\,1,n_v=1\}$ and $\{n_u=1,n_v=1\}$ respectively.
Once fluctuations of the mean field are taken into accout,
we obtain a rather complex structure for
$\sigma_0$ (upper gray peaks). 
As deformation of the cluster sets in, the 
$M=0$ component of the collective dipole
peak corresponding to the spherical cluster ($\eta = 1.0$) gets
fragmented as a consequence of the 
more complex structure of $D^0$ in the resonance region.
The amount of fragmentation depends on the deformation
parameter. The high frequency peak is more fragmented
for $1 < \eta < 1.5$, reaching a rather simple structure with
a dominant peak for $\eta=2$.
The $M=1$ component of the 
photoabsorption cross section shown in Fig.~\ref{NF6} does not present major
new features. The $\sigma_1^{(0)}$ shows a dominant peak for each deformation
$\{n_u=0,n_v=0)$ with $W$ and $B$ orbits contributing in proportion
to the number of valence electrons moving in each of them (see Fig.~\ref{NF3}).
The collective
effects shift the $\sigma_1$  dipole peak to  frequencies below
the one corresponding to the spherical cavity as deformation increases. 

The position of the dominant peaks can be compared with the predictions
of the classical theory. In the oblate case, the $\omega_i$ frequencies
are still given by Eqs.~(\ref{3.3})--(\ref{3.4}) and \cite{17}
\beq
n_z ={{1+e^2}\over{e^3}}\left[\,e-\arctan e\,\right]\:,
\eeq
with  the eccentricity $e$ related to our oblate deformation
parameter by
\begin{equation}
e=\sqrt{\eta^2 -1}\:.
\end{equation}
A simple calculation gives $\omega_{x,y}= 0.912\:\omega_{Mie}$ ,
$~\omega_z = 1.157\:\omega_{Mie}$ for $\eta = 1.5$ and
$\omega_{x,y} = 0.842\:\omega_{Mie}$ ,
$~\omega_z = 1.258\:\omega_{Mie}$ for $\eta = 2$. 
The $M=1$ collective plasmons $(\omega_x,\omega_y)$ are again blue
shifted and in roughly the same amount we observed in the prolate
case ($\sim 7 \%$). However  for $M=0$ the collective peaks
are now blue shifted by $\sim 10 - 15 \%$ 
with respect to the position 
expected from the classical values.

\subsection{Comparison between prolate and oblate clusters}

In this section we compare the most prominent features concerning
the dipole response of spheroidal clusters to an external field. 
We have chosen to show the photoabsorption cross section per valence 
electron ($N = 254$) calculated for a sodium cluster which has been deformed 
to both prolate and oblate shapes. 
We display calculations
corresponding to deformations $\eta=1.0\, (0.25)\, 2$  in both geometries.
In Fig.~\ref{NF7} we plot the corresponding cross sections
calculated both  in the static
mean field (single--particle approximation, shaded black)
and including collective effects (shaded gray). Arrows are plotted as
reference. They place the position of the plasmon peak for a spherical
cluster in the present model. For prolate geometry 
the photoabsorption strenght always displays the characteristic splitting 
into two  pronounced peaks, corresponding
to oscillations along two perpendicular symmetry axes
($M=0$ and $M=1$). The relative
strength of the two collective peaks varies 
with an increasing dominance of the $M=1$ component for
larger deformations. For oblate 
spheroidal geometry instead, an interesting phenomenon occurs: one 
peak dominates the cross section in the $\eta = 1 \div 2$ range.
The high--energy collective plasmon ($M=0$ component)  gets
fragmented and distributed over an interval of frequencies of the order of
$\sim 0.3~\omega_{Mie}$. This fragmentation tends to dissapear at the largest
deformation studied here ($\eta=2$). However, as shown in Fig.~\ref{NF6},
some residual fragmentation 
remains at frequencies similar to the characteristic
frequency of the low--energy plasmon
($M=1$ component). The effect has been traced back to the
appearance of  extra peaks in the "single-particle" $M=0$ strength, that is
related to the existence of two kinds of three-dimensional orbits (W-type
and B-type) in an oblate cavity. It is reasonable to expect that a similar
effect should occur also for more realistic mean fields.

Overall our results give a clear indication on the gross features to be
expected for either prolate or oblate medium size 
sodium clusters. The photoabsorption cross
section is splitted into two peaks for the prolate geometry, but is
mainly dominated by a single peak in the oblate case for the range
of deformations studied here, the smaller peak being further splitted
or fragmented.

\subsection{Scaling properties}

	In this section we investigate the dependence of our results from
the valence electron number $N$ and from the density parameter $r_s$.
By using the explicit expressions of the free propagator $D^0$ given in Sect.II
and in the Appendix it is possible to define
for each geometry (prolate, oblate, spherical), 
at  given cluster deformation $\eta$,
a  corresponding universal propagator $d^{\,0} ({\bf x}',{\bf x},s)$ such that
\beq
\label{sca}
D^{\,0} ({\bf q}',{\bf q},\omega)=N\,r_s{\,^2}\,d^{\,0} 
(R{\bf q}',R{\bf q},\frac{\omega}
{\omega_{_F}})\:,
\eeq
where $\omega_{_F} \equiv \frac{\pi\, p_{_F}}{m\,R}$ 
and $R$ is the cluster radius for spherical geometry.
In Eq.(\ref{sca}) all the dependence from the number of valence electrons
and from the electron density has been extracted from the free propagator $D^{\,0}$
by relating it to a universal function $d^{\,0}$ of the dimensionless parameters
${\bf x} =R{\bf q}$ and $s=\frac{\omega}{\omega_{_F}} $. 
The question we address now
is  whether a similar factorization occurs 
also for the collective propagator $D$.
In general this factorization 
is not possible for an arbitrary interaction
$V$. Indeed, by defining a
function $d({\bf x}',{\bf x},s)$ from the collective propagator $D$ through
Eq.~(\ref{sca}) then the integral equation (\ref{2.1}) leads to the
following integral equation for $d$,
\beq
\label{nosca}
d({\bf x}',{\bf x},s)=d^{\,0} ({\bf x}',{\bf x},s)+\frac{1}{(2\pi)^3\, r_s}
\int d{\bf y}\,d^{\,0}({\bf x}',{\bf y},s)\, V(\frac{y}{R})\,
d({\bf y},{\bf x},s)\:,
\eeq
which explicitly contains the density
parameter $r_s$. An additional $N$ and $r_s$ dependence is implicit
through the $R$ parameter in the force $V$.
Consequently it is not possible to 
determine an obvious  scaling law allowing for the calculation of a universal
collective function $d({\bf x}',{\bf x},s)$. 

In our case $V$ corresponds to a modified Coulomb--type
residual interaction (Eq.~(\ref{force})). Then it is reasonable
to search for additional
properties of Eq.~(\ref{nosca}) in the presence of a pure Coulomb
force. We obtain
\beq
\label{nosca1}
d({\bf x}',{\bf x},s)=d^{\,0} ({\bf x}',{\bf x},s)+\frac{1}{(2\pi)^3}\,
r_s\,N^{\,\frac{2}{3}}
\int d{\bf y}\,d^{\,0}({\bf x}',{\bf y},s)\,\frac{4\,\pi\,e^2}{y^2}\,
d({\bf y},{\bf x},s)\:.
\eeq
In this limiting case, the solution to Eq.~(\ref{nosca1})
depends on the product
$r_s\,N^{\frac{2}{3}}$ only.
Deviations from this result are due to  more realistic
effective interactions 
and they may be expected to be important in the smaller clusters.

In the limit of large $N$ our
microscopic calculations tend to reproduce the results of the macroscopic 
theory, with the plasmon frequency depending only on the electron density, but
not on the total electron number $N$. The $N$-dependence of Eq.(\ref{nosca1})
is not in conflict with this expectation since the dimensionless parameter
$s$ is related to the frequency by an $N$-dependent relation $s\propto
N^{\frac{1}{3}}\, r_s ^2\, \omega$.

\begin{figure}
\caption{
Photoabsorption cross section per valence electron for  prolate
and oblate sodium clusters at several deformations
and for $N=25$ (shaded black) and $N=254$ (shaded gray).}
\label{NF8}
\end{figure}
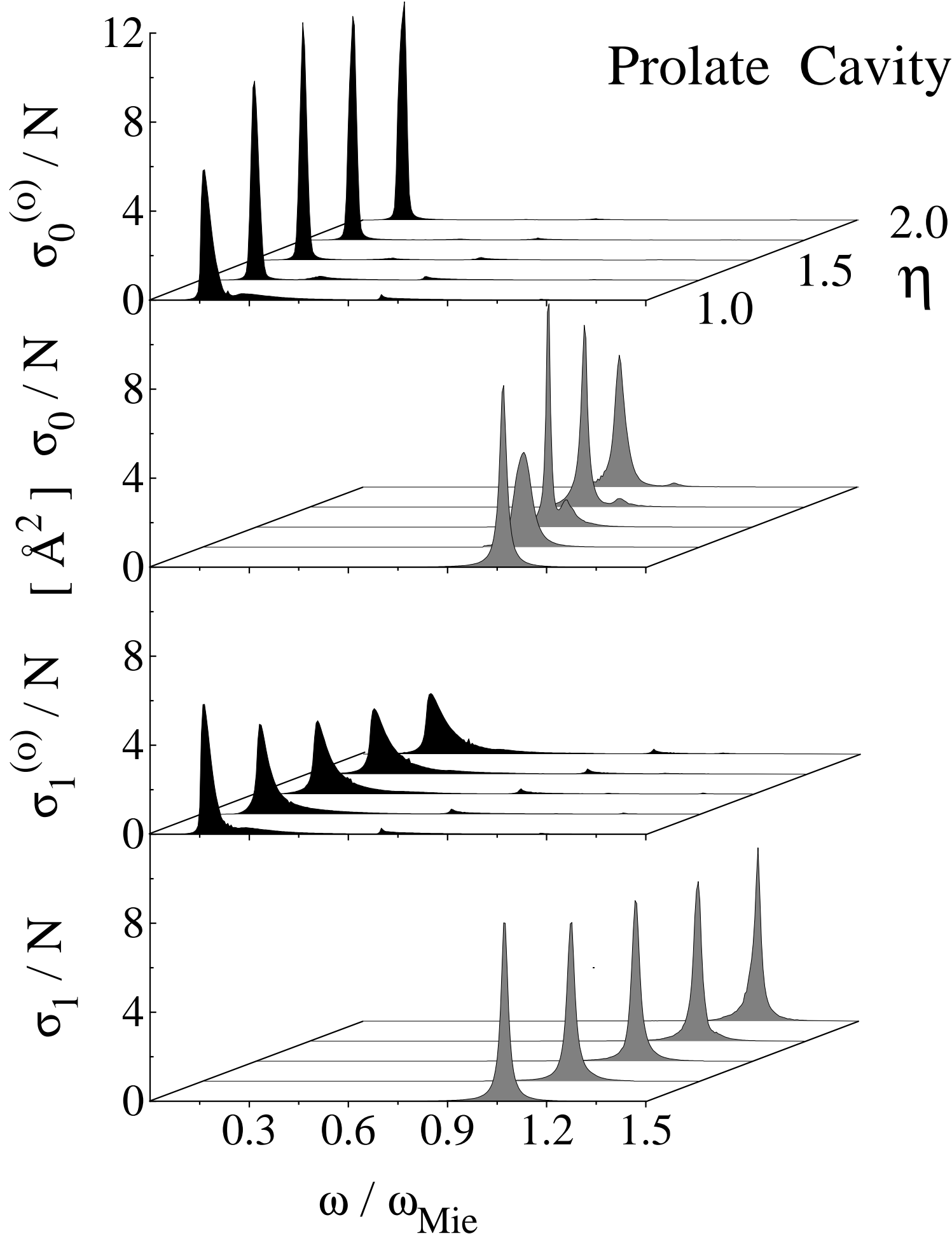
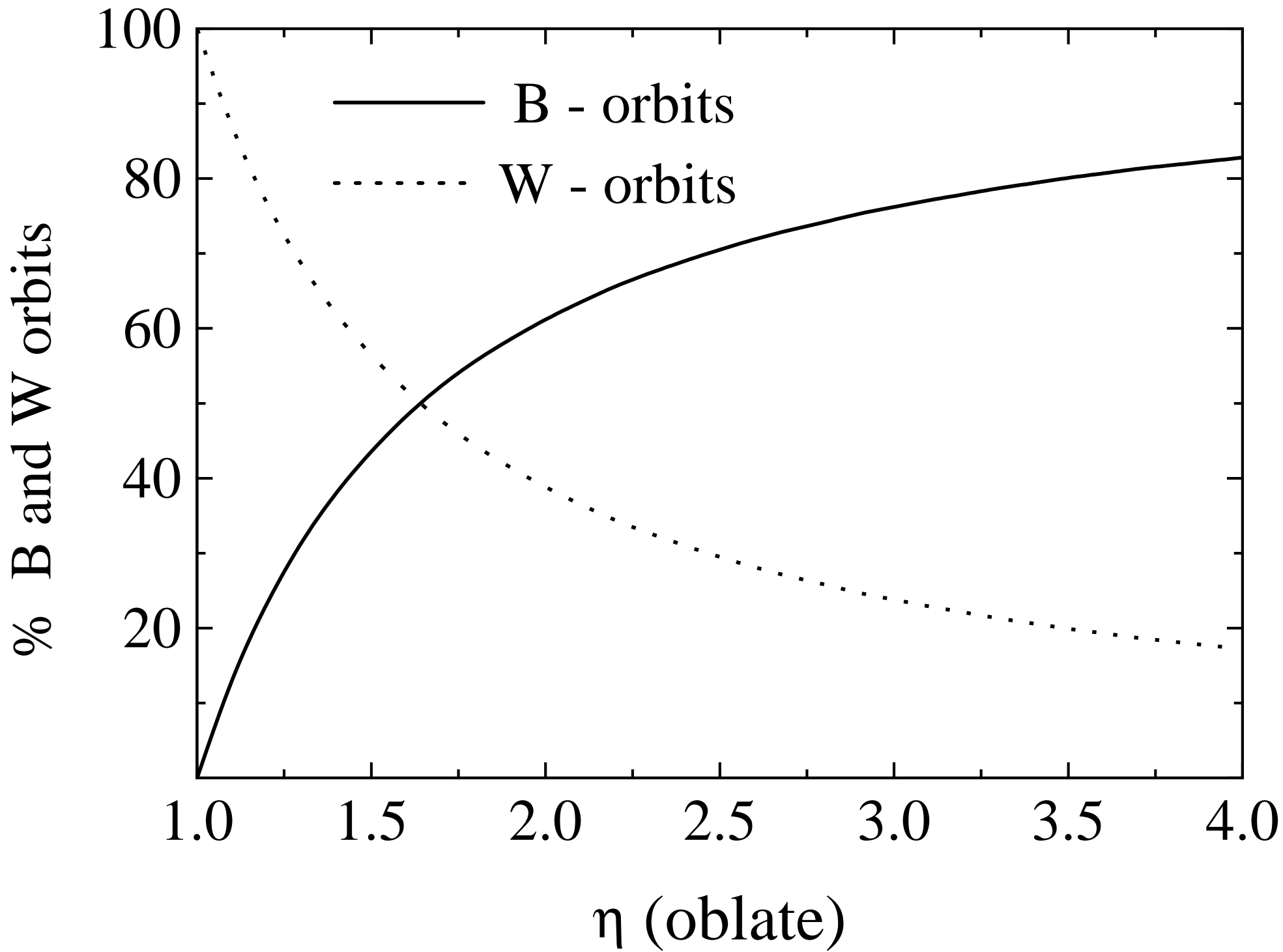
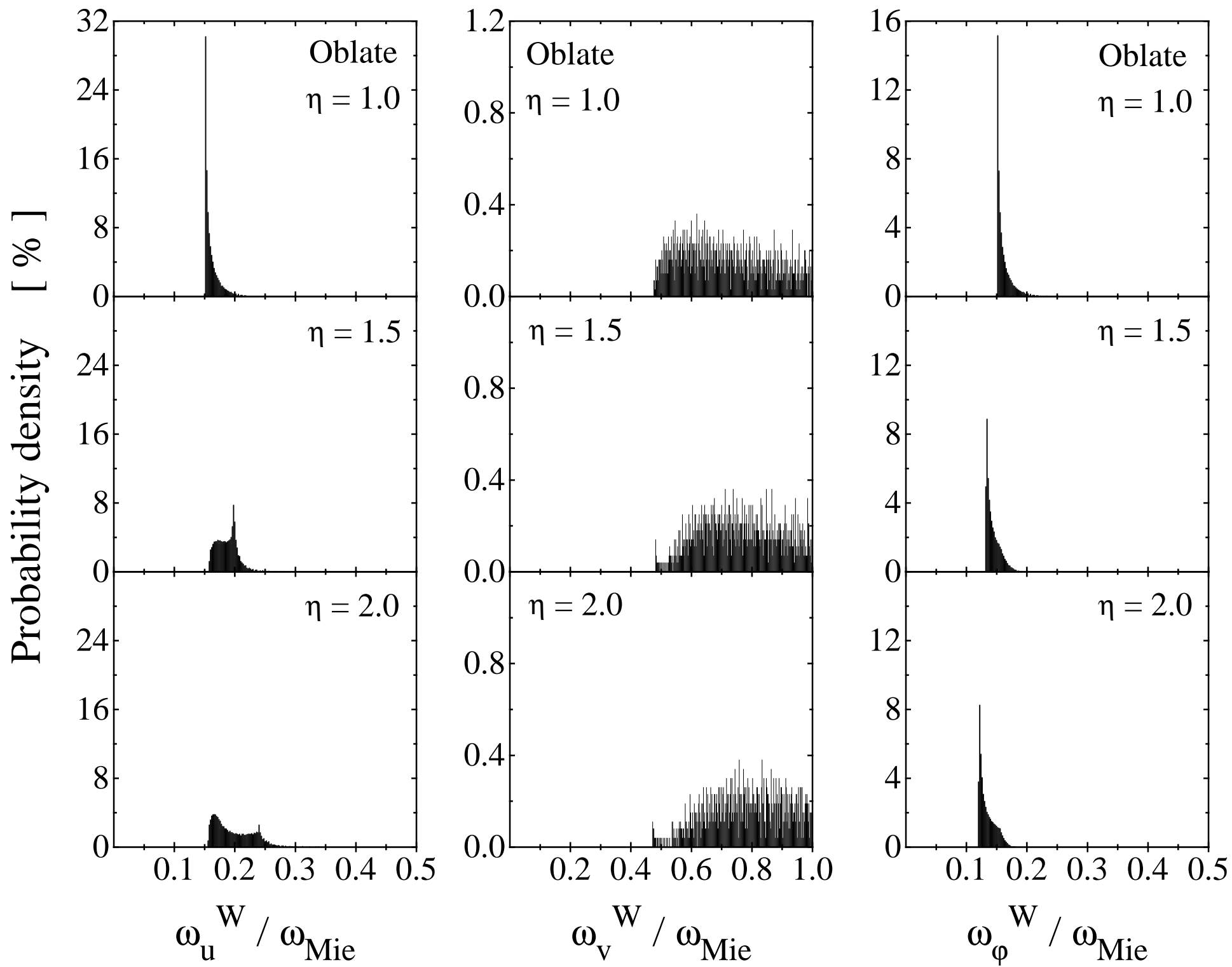
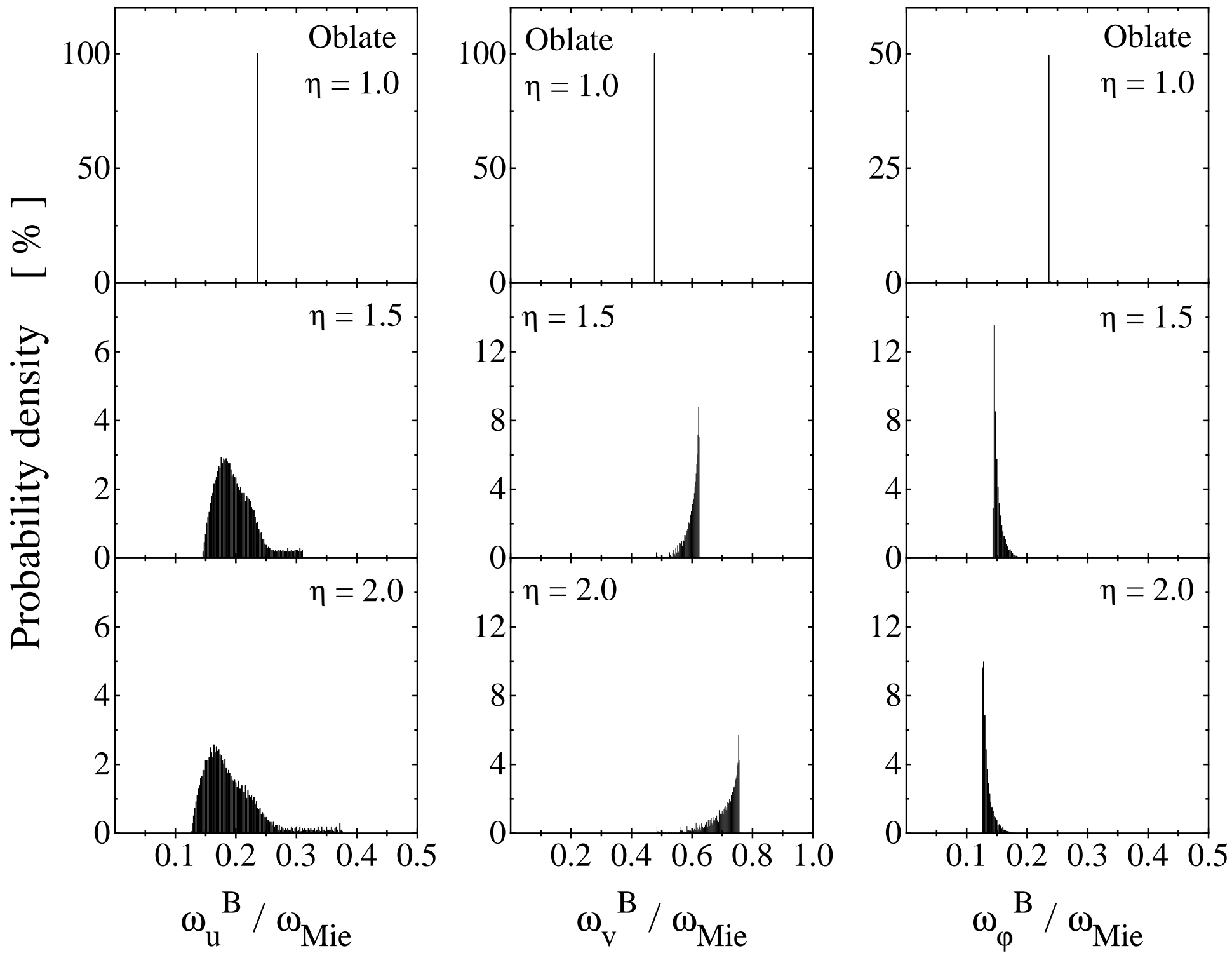
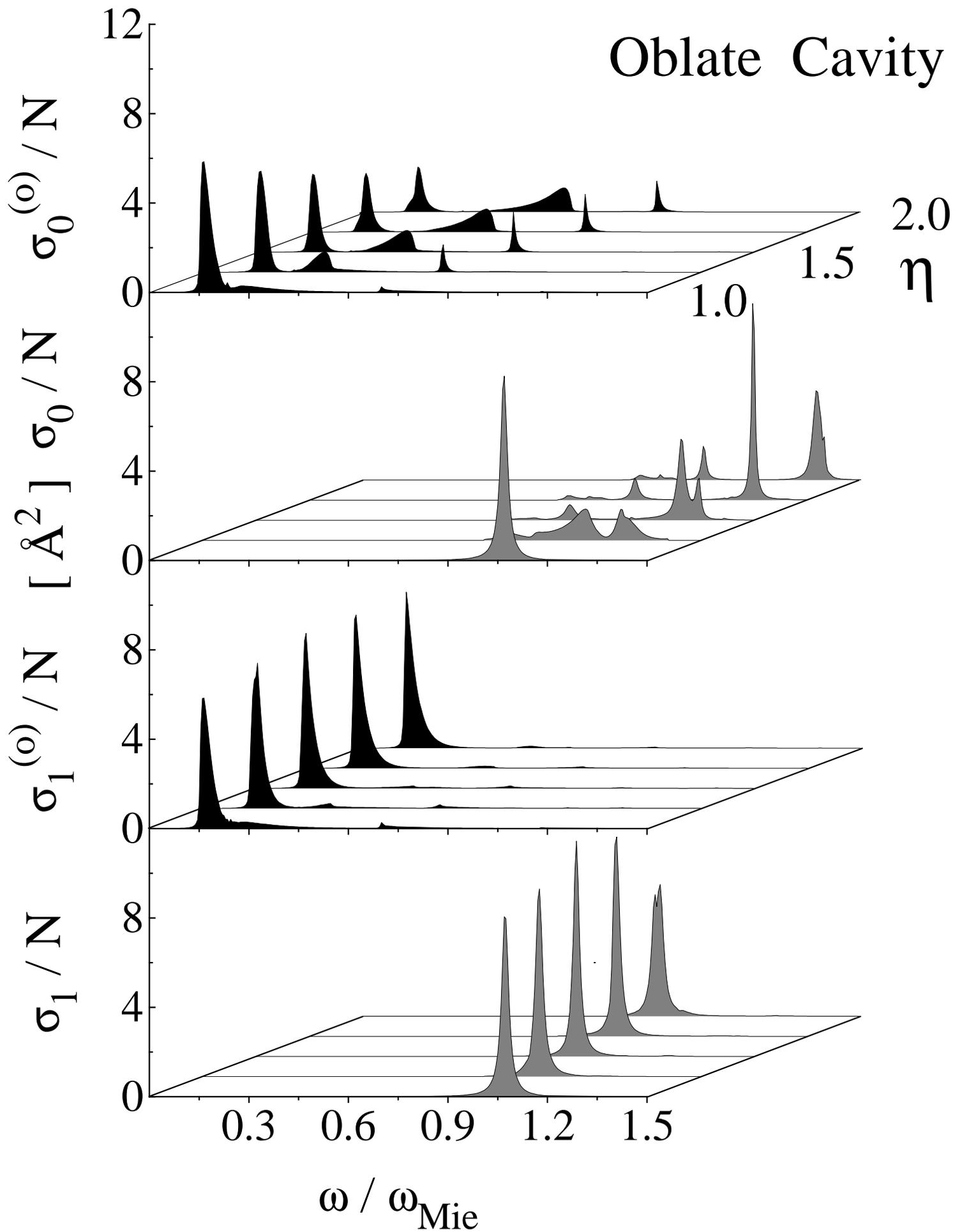
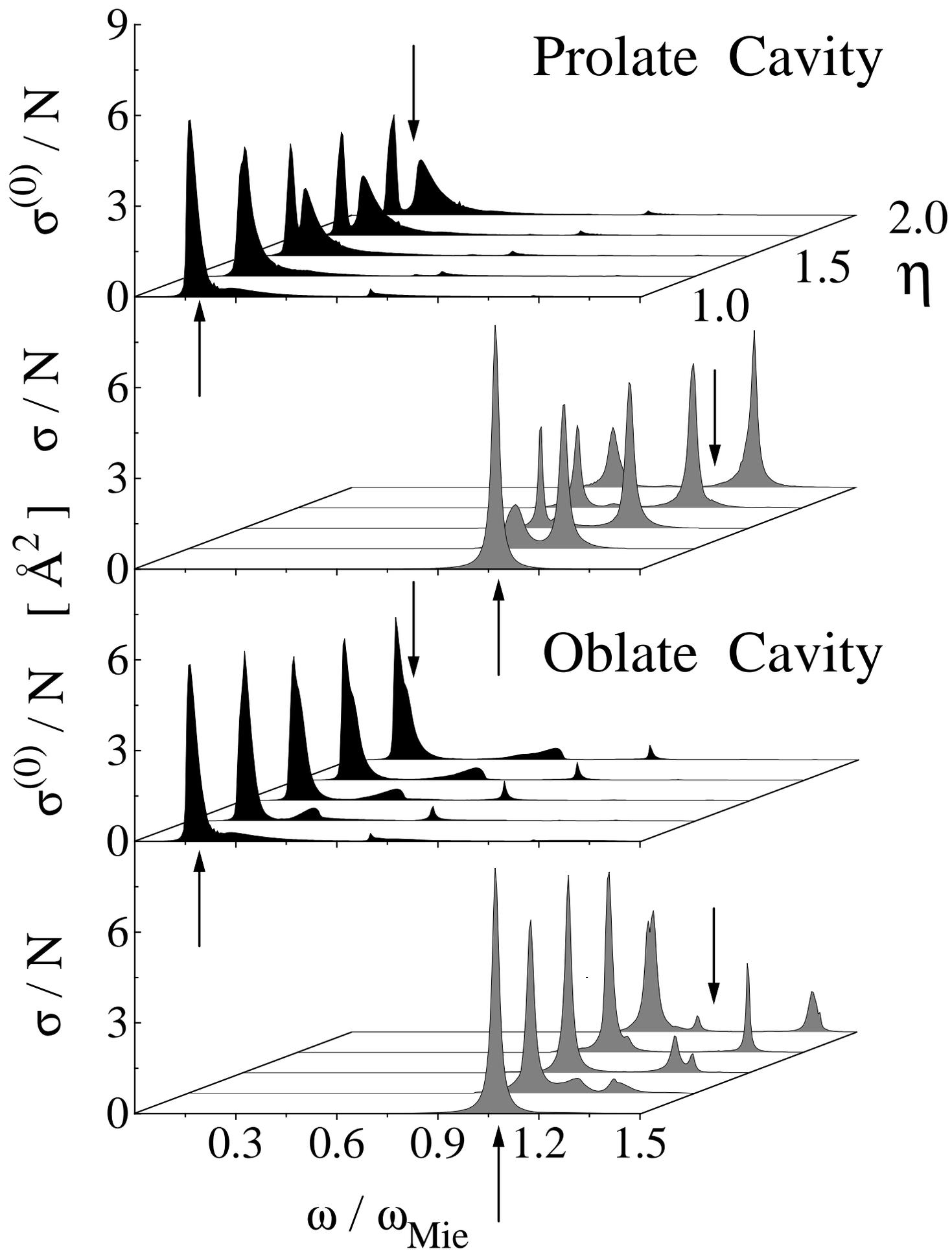
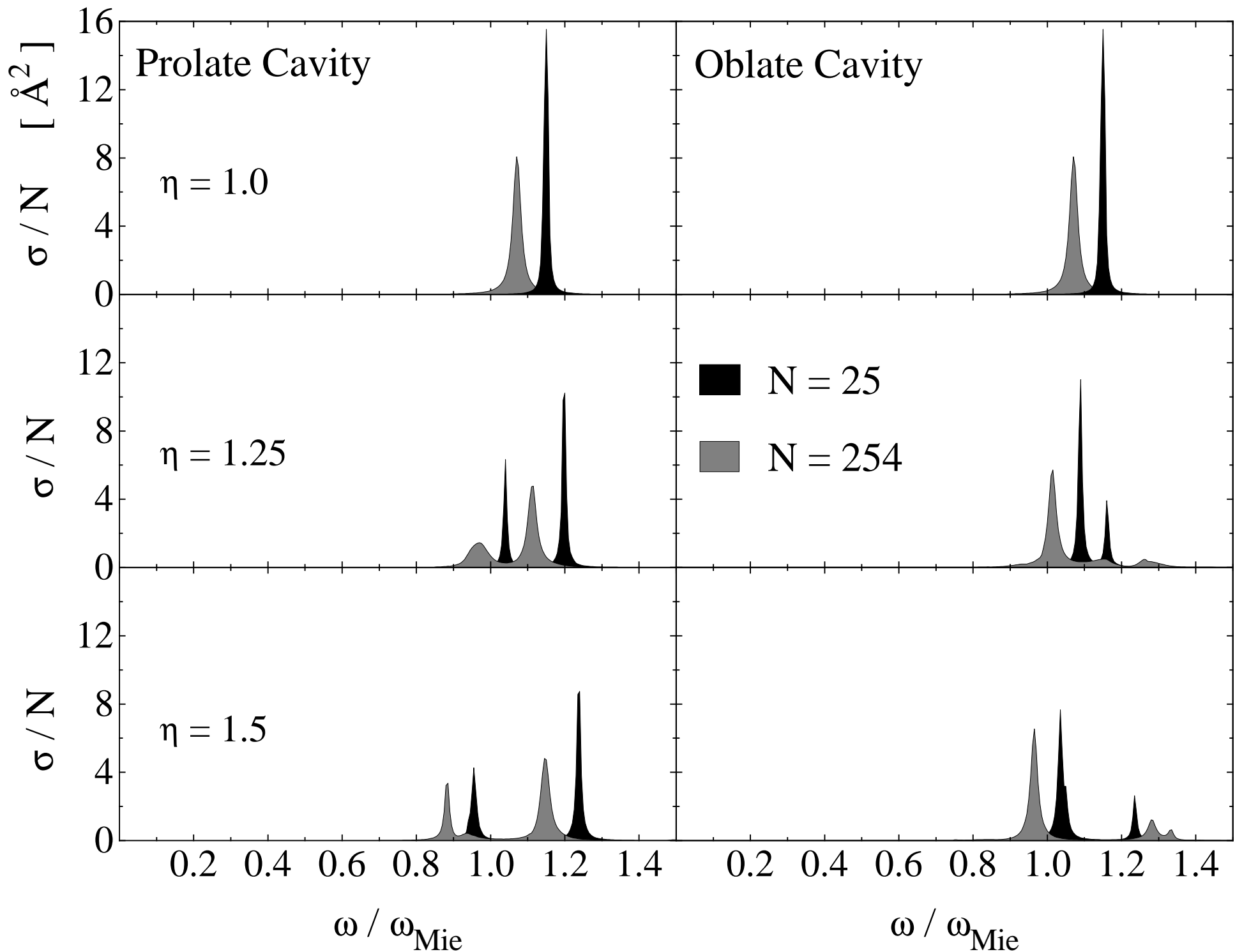

Our discussion indicates that
no exact scaling is to be expected for the photoabsorption cross sections in
the present model.
The actual amount of scaling violation in $D$ does depend on the size,
composition and geometry of the particular cluster considered. 
In order to illustrate
the $N$-dependence of our results we have calculated the collective
response of sodium clusters with 25 atoms, thus changing $N$ by an order of
magnitude with respect to the results reported earlier. 
In Fig.~\ref{NF8} we display the collective response of $N=25$ sodium
clusters (shaded black) together with that for $N=254$ (shaded gray). For
spherical and prolate geometry the $N=25$ plasmons are  blue shifted
by $\sim 10 \%$
with respect to the $N=254$ results, with the overall peak profile showing a
qualitatively similar behaviour in the 
two cases. The light cluster plasmons are
sharper due to a smaller Landau damping. For oblate geometry
we notice that the fragmentation of the high-frequency plasmon found for $N=254$
does not appear for $N=25$. The fragmentation phenomenon is 
$N$-dependent and it disappears for different cluster mass. These
results agree with our general statement about the lack of scaling with $N$
of the collective propagator $D$. 

The study of clusters with a larger number of valence electrons
($N > 254$) will not provide much new information since the
theory starts approaching the classical limits. We have confirmed this
by doing calculations for sodium clusters with $N = 2500$.
We observe a red shift with respect to the $N=254$ case of the order
of $3 \%$ and no fragmentation of the $M=0$ peak.

We have investigated also the electron-density dependence of our results.
Calculations of the photoabsortion cross section for
prolate and oblate
sodium and potassium clusters ($r_s = 5$) at  $N=254$ 
show very similar peak profiles (including fragmentation)
when the frequency axis is rescaled with the appropriate value 
of $\omega_{Mie}$. Although the density of valence electrons
was decreased by about a factor 2, these small variations
can be understood on the basis of Eq.~(\ref{nosca1}) where the
relevant parameter ($r_s\,N^{\frac{2}{3}}$) was changed by $25 \%$.

\section{Summary and conclusions}

In this paper a semiclassical theory of linear response 
based on the Vlasov equation \cite{7,8} has been extended to 
spheroidal systems and then applied to study the peak profile of 
surface plasmon resonances in medium size deformed atomic clusters.
Assuming a spheroidal cavity model to describe both prolate and
oblate clusters we have been able to calculate the gross features of
the cluster response to an external field of frequency $\omega$ 
as a function of the cluster deformation. 

Two main general results emerge from our calculations. One of them refers 
to the splitting of the collective dipole peak with increasing 
deformation and to the position of these peaks. On this we have commented
at lenght in Sec. III.  The other main result is related 
to the width of the dipole peaks. In the present model the single--particle
dipole resonance does display a width which is due to the non linearity of the
assumed equilibrium mean field. This single-particle width generates a width 
in the collective plasmon resonances through a mechanism that is analogous
to the Landau damping in homogeneous systems. Our calculated 
width is not sufficient to reproduce the observed plasmon width.
However our width is underestimated
since we have included  only the first few frequency modes 
when evaluating $D^0$ in Eq.~(\ref{2.4}). The neglected terms would 
increase the imaginary part of $D^0$ in the region of $\omega 
\simeq \omega_{Mie}$ and thus increasing the Landau damping. Estimates based on
numerical calculations including more modes
set this effect at about $20 \%$, which is far from sufficient to explain
the observed values.  Thus
more sophisticated effects, like the possible coupling to surface
vibrations \cite{18}, should perhaps be taken into account.

The fragmentation of the high--energy plasmon peak for medium size
oblate clusters
is a definite prediction of the present theory and reflects the existence
of nontrivial dynamics. The relative importance of this 
effect  depends on several elements such as the shape and size of
the cluster and  the strength of the effective interaction.

The normalized cluster dipole response shows no simple scaling properties
with $N$  at given density ($r_s = 4$). The position of the plasmon
peaks is weakly dependent on $N$, changing by $\sim 10\%$ in the 
$25 < N < 2500$ mass region while fragmentation of the high energy
component dissapears for light ($N \sim 25$) and heavy ($N \sim 2500$)
oblate clusters. On the other hand, no major changes in the photoabsorption
profile are observed when  density decreases by  a factor $\sim 2$
($r_s = 5$) for a given number $N$ of valence electrons.

Extensions to this work are clearly welcomed. In particular, a close
comparison to experiment is required to asses in detail some of the
physics missing in the present model. Nevertheless we have established a solid
framework to  classically understand the gross features of
complicated, intrinsically quantum systems.

\section*{Acknowledgments}

A.D. is grateful to Prof. D. M. Brink for useful suggestions.
F.A.B. thanks support from FONDECYT grant 1960690, Fundaci\'on Andes,
INFN and Ministero della Pubblica Istruzione, Italy.

\appendix
\section*{}

In this Appendix we specify the details for the expressions given in
Sect. II. The propagator $D^{0}$ is determined entirely by the 
single-particle motion in the equilibrium mean field. The classical 
motion of a point particle in a spheroidal cavity with perfectly
reflecting walls has been studied in Ref.~\cite{5} for prolate cavities 
and in Ref.~\cite{6} for both prolate and oblate cavities (see also 
Refs.~\cite{19,20}~). The authors of Ref.~\cite{6} have pointed out
that the three-dimensional motion in a prolate cavity is simpler than
that in an oblate cavity. We give first a detailed description for prolate
geometry. Then, through a simple transformation, the corresponding results
for the oblate shape can be recovered. Furthermore we remark the non trivial 
differences related to particle motion in the two geometries. Our notation 
follows closely that of Ref.~\cite{5}.

\subsection{Prolate cavity}

To describe the prolate shape, let us introduce the prolate spheroidal
coordinates $\{u,v,\varphi \}$ through their relation to the cartesian
coordinates,
\begin{eqnarray}
\label{B.1}
x & = & \xi_{_{P}} \cos u \sinh v \cos\varphi \:, \qquad
-\,{\textstyle \frac{\pi}{2}} \leq  u \leq {\textstyle \frac{\pi}{2}} \:,
\nonumber \\
y & = & \xi_{_{P}} \cos u \sinh v \sin\varphi \:, \qquad \quad \;
0 \leq  v \leq \infty \:,  \nonumber \\
z & = & \xi_{_{P}} \sin u \cosh v \:, \qquad \qquad \quad \;\:
0 \leq  \varphi \leq 2\pi \:,
\end{eqnarray}
with
\begin{equation}
\label{B.2}
\xi_{_{P}} = \sqrt{c^2 -a^2} \:.
\end{equation}
In the last equation $c$ and $a$ are the larger and smaller semiaxes 
respectively. The two focal points are at $z=\pm\,\xi_{_{P}}$.  

The equilibrium Hamiltonian $h_0$ can be expressed in terms of the
spheroidal coordinates $\{u,v,\varphi\}$ and their conjugate momenta 
$\{p_u,p_v,p_{\varphi}\}$ \cite{5,6}. Assuming that particles move in a 
static cavity of spheroidal shape,  represented
by a potential energy ${\mathcal{V}}(v)$, then $h_0$ is  given by
\begin{equation}
\label{B.3}
{h_0}^{(P)} (u,v,\varphi,p_u,p_v,p_\varphi)  = 
\frac{ p_{u}^2 +p_{v}^2 }{ 2\,m\,\xi_{_{P}}^2\,
(\cosh^2 v - \sin^2 u) } +
\frac{ p_{\varphi}^2 }{ 2\,m\,\xi_{_{P}}^2\, \sinh^2 v \cos^2 u} + 
{\mathcal{V}}(v) \:,
\end{equation}
for the prolate configuration. The potential energy for sharp walls is
\begin{eqnarray}
\label{B.4}
{\mathcal{V}}(v) & = & 0 \qquad \qquad ~v < v_2^P \:, \nonumber \\
     & = & \infty \qquad \qquad v \ge v_2^P \:,
\end{eqnarray}
and the $v_2^P$ parameter is determined by the shape of the cavity,
\begin{equation}
\label{B.5}
\sinh v_2^P = \frac{a}{\xi_{_{P}}} \:.
\end{equation}

The Hamiltonian (Eq.~(\ref{B.3})) is integrable and the 
particle motion could, in principle, be described in terms of 
the angle variables $\{\Phi_u,\Phi_v,\Phi_{\varphi}\}$ and of the
three conjugate action integrals
\begin{equation}
I_u =  \frac{1}{2 \pi}\,\oint{p_u du} \:, \qquad
I_v =  \frac{1}{2 \pi}\,\oint{p_v dv} \:, \qquad
I_{\varphi} = \frac{1}{2 \pi}\,\oint{p_{\varphi} d\varphi} \:. 
\end{equation}
However, in the spirit of Ref.~\cite{7}, the following three other 
constants of the motion $\{E,\epsilon,{\lambda}_z\}$  can be conveniently 
used instead of the action integrals. These new constants are:
the particle energy $E$, the separation variable $\epsilon$ and
the $z$-component of the particle angular momentum $\lambda_z$ 
(which coincides with the action variable $I_{\varphi}$ and with
the generalized momentum $p_\varphi$). The constant of motion $\epsilon$ 
plays a role analogous to the magnitude of the particle angular momentum 
in the spherical case.

With the help of the Vlasov equation the angle variables can be 
explicitly expressed in terms of these three constants of the
motion $\{E,\epsilon,{\lambda}_z\}$ and of the spheroidal coordinates 
$\{u,v,\varphi\}$. The derivation, based on separation of variables 
in the linearized Vlasov equation, is lengthy but straightforward, 
hence we do not report all the details here \cite{note}. 

The generalized momenta $p_{u,v}$ are
\begin{eqnarray}
\label{B.6}
p_u&=&\xi_{_{P}}\,\sqrt{2mE}\,\sqrt{\sigma_1^P - U_{_P}(u,\sigma_2^P)} \:,
\nonumber \\
p_v&=&\xi_{_{P}}\,\sqrt{2mE}\,\sqrt{V_{_P}(v,\sigma_2^P)-\sigma_1^P}\:.
\end{eqnarray}
The two dimensionless constants of the motion $\sigma_1^P$ and $\sigma_2^P$
are  defined as in Ref.~\cite{5},
\begin{equation}
\label{B.8}
\sigma_1^P  =  \frac{\epsilon}{E} \:, \qquad
\sigma_2^P  =  \frac{\lambda_{z}^2}{2\,m\,E\, \xi_{_{P}}^2} \:,
\end{equation}
and the effective potentials in Eq.~(\ref{B.6}) for the $u$ and $v$ 
coordinates are given by
\begin{eqnarray}
\label{B.9}
U_{_P}(u,\sigma_2^P) & = & \sin^2 u + \frac{\sigma_2^P}{\cos^2 u} \:,
\nonumber \\
V_{_P}(v,\sigma_2^P) & = & \cosh^2 v - \frac{\sigma_2^P}{\sinh^2 v}\:.
\end{eqnarray}

Particles on three--dimensional orbits move between two confocal 
ellipsoids with $v_1^P \le v \le v_2^P$ and 
$u_1^P \le u \le u_2^P$, with $v_{1,2}^P$ and $u_{1,2}^P$ the turning
points obtained as solutions to $p_v =0$ and $p_u =0$ respectively 
(see Ref.~\cite{5} for details). Introducing the quantities
\begin{equation}
\label{B.10}
t_{\pm}^P = \sqrt{\left(\frac{1 - \sigma_1^P}{2}\right) \pm
\sqrt{\left(\frac{1 - \sigma_1^P}{2}\right)^2 + \sigma_2^P}} \:,
\end{equation}
these turning points are obtained from
\begin{eqnarray}
\label{B.11}
\cosh(v_1^P) & = & \sqrt{1 - (t_{-}^P)^2\,} \:, \nonumber \\
\cos(u_2^P) & = & t_{+}^P\:, \qquad \qquad u_1^P  =  -\,u_2^P \:,
\end{eqnarray}
and $v_2^P$ from Eq.~(\ref{B.5}). 

It is our purpose here to specify the integration limits in
Eq.~(\ref{dzero}), and to derive explicit expressions for the 
eigenfrequencies $\omega_{u,v,\varphi}$ as well as for the Fourier 
coefficients $Q^{(LM)}_{n_u,n_v,n_\varphi}$ appearing in the 
expression for $D^0$.

To obtain the integration range for the constant of motion $\epsilon$ 
we realize that the turning point $u_2^P$  exists only if 
$\cos u_2^P \le 1$ in Eq.~(\ref{B.11}). Also the presence of an infinite 
potential barrier at the surface of the cavity implies $p_v(v_2^P) = 0$. 
These two conditions impose  constraints on $\sigma_1^P$,
\begin{eqnarray}
\label{B.12}
{\left(\,\sigma_1^P\,\right)}_{min} &=& \sigma_2^P \:, \nonumber \\
{\left(\,\sigma_1^P\,\right)}_{max} &=& \cosh^2 v_2^P - \frac{\sigma_2^P}
{\sinh^2 v_2^P} \:.
\end{eqnarray}
These constraints determine the integration range for $\epsilon$. Thus,
\begin{equation}
\label{B.13}
\int d\epsilon \quad \longrightarrow \quad \int_{\epsilon_{-}}^{\epsilon_{+}}
d \epsilon \:,
\end{equation}
with
\begin{equation}
\label{B.14}
\epsilon_{-} = \frac{{\lambda}_z^2}{2\, m \,\xi_{_{P}}^2} \:, \qquad
\epsilon_{+} = E_F\,\cosh^2 v_2^P - \frac{\epsilon_{-}}{\sinh^2 v_2^P} \:.
\end{equation}

The integration range for the constant of motion $\lambda_z$ is determined
from the possible values of the particle angular momentum along the
symmetry axis. Thus,
\begin{equation}
\label{B.15}
\int d\lambda_z \quad \longrightarrow \quad \int_{\lambda_{-}}^{\lambda_{+}}
d \lambda_z \:,
\end{equation}
where 
\begin{equation}
\label{B.16}
\lambda_{\pm} = \pm\,(p_F a) \:,
\end{equation}
with $p_F$ the Fermi momentum associated to the Fermi energy $E_F$.

The evaluation of the Fourier coefficients in Eq.~(\ref{fou})  can be 
simplified by noticing that
\begin{equation}
\label{B.23}
n_u\Phi_u+n_v\Phi_v+n_\varphi \tilde\Phi_\varphi=s_{\bf n}(u)+s'_{\bf n}(v)\:.
\end{equation}
The phases $s_{\bf n}(u)$ and $s'_{\bf n}(v)$ are given by
\begin{eqnarray}
\label{B.24}
s_{\bf n}(u)&=&{\bf n}\cdot \bomeg\, \tau_u(u)+{\bf n}\cdot \bbomeg\,
\alpha_u(u) -n_{\varphi}\gamma_u(u)\:, \nonumber \\
s'_{\bf n}(v)&=&{\bf n}\cdot \bomeg\,\tau_v(v)+{\bf n}\cdot \bbomeg\,
\alpha_v(v) - n_{\varphi}\gamma_v(v)\:.
\end{eqnarray}
with $\bomeg$ the frequency vector defined in Eq.~(\ref{omegas}), 
$\bbomeg$ an auxiliary frequency vector
and the auxiliary functions $\{\,\alpha_{u,v},\tau_{u,v},\gamma_{u,v}\,\}$
given by
\begin{eqnarray}
\label{B.19}
\alpha_u(u)& = &m\, \xi_{_{P}}^2 \int_{u_{min}}^u \frac{d u\,'}{p_u(u\,')} \:,
\qquad \qquad \qquad \: \:
\alpha_v(v) = -\, m\, \xi_{_{P}}^2 \int_{v_{min}}^{v} \frac{d v\,'}{p_v(v\,')}
\:,\nonumber \\
\tau_u(u)& =& -\,m\, \xi_{_{P}}^2 \int_{u_{min}}^u \frac{\sin^2{ u\,'}}
{p_u(u\,')} d u\,' \:,
\qquad \qquad
\tau_v(v) =  m\, \xi_{_{P}}^2 \int_{v_{min}}^{v} \frac{\cosh^2{v\,'}}
{p_v(v\,')} d v\,' \:,\nonumber \\
\gamma_u(u) &=& \lambda_z \int_{u_{min}}^u \frac{d u\,'}
{\cos^2 u\,'p_u(u\,')} \:,
\qquad \qquad \:\:
\gamma_v(v) =\lambda_z \int_{v_{min}}^{v} \frac{d v\,'}
{\sinh^2 v\,'p_v(v\,')} \:. 
\end{eqnarray}
In these equations the upper integration limit is a variable while
the lower one is the corresponding inner turning point for the geometry under 
study. For the prolate case $u_{min} = u_1^P$  and  $v_{min}=v_1^P$. All the
integrals appearing in Eq.~(\ref{B.19}) can be easily expressed
in terms of elliptic integrals and evaluated numerically.

The natural frequencies of the unperturbed trajectories are the
three components of the frequency vector $\bomeg$ (Eq.~(\ref{omegas})).
They can be expressed in terms of these integrals as
\begin{eqnarray}
\label{B.21}
\omega_u (E,\epsilon,\lambda_z )&=& \pi\, \frac{-\,\alpha_v (v_2^P )}
{\left[\,\alpha_u (u_2^P )\tau_v (v_2^P)-
\alpha_v (v_2^P) \tau_u (u_2^P )\,\right]}\:,
\nonumber \\
\omega_v (E,\epsilon,\lambda_z )&=& \pi\, \frac{\alpha_u (u_2^P )}
{\left[\,\alpha_u (u_2^P )\tau_v (v_2^P)-
\alpha_v (v_2^P) \tau_u (u_2^P )\,\right]}\:,
\nonumber \\
\omega_\varphi (E,\epsilon,\lambda_z )&=&\frac{1}{\pi}\left[\,\omega_v 
\,\gamma_v (v_2^P ) + \omega_u\, \gamma_u (u_2^P )\,\right]\:.
\end{eqnarray}
Correspondingly, the auxiliary frequency vector $\bbomeg$ has components
\begin{eqnarray}
\label{B.22}
\Omega_u(E,\epsilon.\lambda_z )&=&-\, \omega_u\,\frac{\tau_v (v_2^P )}
{\alpha _v (v_2^P )}\:, \nonumber \\
\Omega_v(E,\epsilon.\lambda_z )&=&-\, \omega_v\,\frac{\tau_u (u_2^P )}
{\alpha _u (u_2^P )}\:, \nonumber \\
\Omega_{\varphi}(E,\epsilon,\lambda_z )&=&\frac{1}{\pi}\left[\,\Omega_v 
\,\gamma_v (v_2^P )+\Omega_u\, \gamma_u (u_2^P )\,\right]\:.
\end{eqnarray}

Note that from the expressions given above we can easily obtain
the explicit form of the angle variables in terms of the spheroidal 
coordinates. For example the explicit form of $\Phi_u (u,v)$ can be  
derived from  Eq.~(\ref{B.23}) by setting $n_u =1$, $n_v =0$, 
$n_{\varphi} =0$. The angle variable $\Phi_\varphi$ takes the form
$\Phi_\varphi=\tilde\Phi_\varphi (u,v)+\varphi$.

The evaluation of the Fourier coefficients in Eq.~(\ref{fou}) is made
explicit by using the identity
\begin{eqnarray}
\label{B.27}
\oint du\oint dv \left|\frac{\partial (\Phi_u,\Phi_v)}
{\partial (u,v)}\right|\, e^{-i(n_u\Phi_u+n_v\Phi_v+n_\varphi
\tilde\Phi_\varphi)}&=& \nonumber \\
4\int_{u_1^P}^{u_2^P}du\int_{v_1^P}^{v_2^P}dv
\left|\frac{\partial (\Phi_u,\Phi_v)}
{\partial (u,v)}\right|\,\cos[s_{\bf n}(u)]\cos[s'_{\bf n}(v)]\:,& &
\end{eqnarray}
and evaluating the Jacobian,
\begin{equation}
\left|\frac{\partial (\Phi_u,\Phi_v)}{\partial (u,v)}\right| =
\left(\,m\, \xi_{_{P}}^2\,\right)^2\,
\frac{\omega_u\,\omega_v}{p_u\, p_v} \left[\,\frac{\tau_u(u_2^P)}
{\alpha_u(u_2^P)} - \frac{\tau_v(v_2^P)}{\alpha_v(v_2^P)}\,\right]\,
\left(\cosh^2 v -\sin^2 u\right).
\end{equation}
Then,
\begin{equation}
\label{B.29}
Q^{(LM)}_{n_u,n_v,n_{\varphi}}=\delta_{M,n_{\varphi}}
\int_{u_1^P}^{u_2^P} {du\over{p_u}}\int_{v_1^P}^{v_2^P} {dv\over{p_v}}\,F(u,v)
\cos[s_{\bf n}(u)]\cos[s'_{\bf n}(v)] \:,
\end{equation}
with
\begin{eqnarray}
\label{B.29x}
F(u,v)& = & i^L \frac{4}{\pi}\,\left(\,m\, \xi_{_P}^2\,\right)^2\,
\omega_u \, \omega_v 
\left[\,\frac{\tau_u (u_2^P )}{\alpha_u (u_2^P )}- \frac{\tau_v (v_2^P )}
{\alpha_v (v_2^P )}\,\right]
\left[\,\cosh^2 v-\sin^2u\,\right] \nonumber \\
  & & \times \,Y_{LM}(\theta(u,v),0)\,j_L (qr(u,v))\:.
\end{eqnarray}
The radial coordinate $r$ and the polar angle $\theta$ can be easily
expressed in terms of the $(u,v)$ variables by using the relations (\ref{B.1}).

The present calculation is simpler if parity selection rules
are taken into account. These selection rules originate from the fact
that the effective potential $U_{_P}(u,\sigma_2^P)$ is an even function of
$u$ and as a direct consequence of the spheroidal geometry invariance under
the reflection $z\to -z$. Because of this symmetry some of the Fourier 
coefficients (Eq.~(\ref{fou})) vanish. This can be proved by using the
following relations,
\begin{eqnarray}
\label{B.30}
s_{\bf n}(-u) & = & n_u \pi -s_{\bf n}(u) \:, \nonumber \\
\cos[s_{\bf n}(-u)] & = & (-)^{n_u}\cos[s_{\bf n}(u)]\:, \nonumber \\
\cos[\theta(-u,v)] & = & -\cos[\theta(u,v)] \:\:\:=\:\:\:
\cos[\pi-\theta(u,v)]\:,\nonumber \\
Y_{LM}(\theta(-u,v),0)& = & (-)^{L-M}Y_{LM}(\theta(u,v),0)\:.
\end{eqnarray}
All the remaining factors in (\ref{B.29}) are even functions of $u$. Then
\begin{equation}
\label{B.31}
Q^{(LM)}_{n_u,n_v,n_{\varphi}}=\delta_{M,n_{\varphi}}[1+(-)^{n_u +L-M}]
\int_{u_0}^{u_2^P} {du\over{p_u}}\int_{v_1^P}^{v_2^P} {dv\over{p_v}}\,F(u,v)
\cos[s_{\bf n}(u)]\cos[s'_{\bf n}(v)] \:,
\end{equation}
with $u_0 = 0$. Equation~(\ref{B.31}) implies that, 
for example, for $L=1$ and $M=1$ we need to sum only over {\it even} values 
of $n_u$, while for $M=0$ we must take only {\it odd} values of $n_u$.
It means also that $D^{0}_{L'LM}=0$ unless $(-)^{L'}=(-)^L$.

\subsection{Oblate cavity}

The transformation analogous to (\ref{B.1}) for oblate geometry
can be obtained from those equations by exploiting the identities
\beq
\sinh(v \pm i\,{\textstyle \frac{\pi}{2}})=\pm\, i\cosh (v)\:, \qquad 
\cosh(v \pm i\,{\textstyle \frac{\pi}{2}} )=\pm\, i\sinh (v)\: .
\eeq
Then the following formal replacements should be made in (\ref{B.1})
to obtain now the relation between the oblate spheroidal
coordinates $\{u,v,\varphi \}$ and the cartesian coordinates,
\begin{equation}
\label{B.32}
\xi_{_P} \longrightarrow i\,\xi_{_O}\:, \qquad \qquad
v \longrightarrow v - i\,{\textstyle \frac{\pi}{2}} \:,
\end{equation}
where
\begin{equation}
\label{B.33}
\xi_{_O} =\sqrt{a^2 - c^2}\:,
\end{equation}
is the radius of the focal circle and $a$ is now the largest semiaxis.
Applying the same transformation to the kinetic enegy part in Eq.~(\ref{B.3})
we obtain the corresponding equilibrium Hamiltonian for an oblate cavity. The
potential energy is still given by Eq.~(\ref{B.4}), with $v_2^P$ being 
replaced by $v_2^{\,O}$,
\begin{equation}
\label{B.35}
\cosh v_2^{\,O} = \frac{a}{\xi_{_{O}}} \:.
\end{equation}
Note that this relation can be obtained by applying the transformation
(\ref{B.32}) to Eq.~(\ref{B.5}).

The generalized momenta $p_{u,v}$ are obtained by applying the
transformation (\ref{B.32}) to the expressions in (\ref{B.6}) and by 
making the replacements
\begin{equation}
\label{B.36}
\sigma_1^{\,O} = 1 - \sigma_1^{P}\:, \qquad \qquad
\sigma_2^{\,O} =  \frac{\lambda_{z}^2}{2\,m\,E\, \xi_{_{O}}^2} =
-\,\sigma_2^{P}\:.
\end{equation}
The generalized momenta for the oblate cavity are then 
expressed as
\begin{eqnarray}
\label{B.37}
p_u&=&\xi_{_{O}}\,\sqrt{2mE}\,\sqrt{\sigma_1^{\,O} - U_{_O}(u,\sigma_2^{\,O})} 
\:,\nonumber \\
p_v&=&\xi_{_{O}}\,\sqrt{2mE}\,\sqrt{V_{_O}(v,\sigma_2^{\,O})-\sigma_1^{\,O}}\:,
\end{eqnarray}
with the oblate effective potentials
\begin{eqnarray}
\label{B.38}
U_{_O}(u,\sigma_2^{\,O}) & = & \cos^2 u + \frac{\sigma_2^{\,O}}{\cos^2 u} \:,
\nonumber \\
V_{_O}(v,\sigma_2^{\,O}) & = & \cosh^2 v + \frac{\sigma_2^{\,O}}{\cosh^2 v}\:.
\end{eqnarray}
Of course we still have $p_{\varphi}=\lambda_z$.

In Ref.~\cite{6} it has been pointed out that the effective potentials 
$U_{_O}(u,\sigma_2^{\,O})$ and $V_{_O}(v,\sigma_2^{\,O})$ for oblate 
geometry may exhibit a non monotonic behaviour for some range of values of 
the parameter $\sigma_2^{\,O}$. As a consequence the phase space of an 
oblate spheroidal cavity is divided into two parts and there are two kinds 
of three-dimensional orbits (plus a separatrix that, however, has zero weight
in our calculations). For  their description we introduce the 
equivalent to the quantities $t_{\pm}^P$ in Eq.~(\ref{B.10}) for the oblate 
case,
\begin{equation}
\label{B.39}
t_{\pm}^{\,O} = \sqrt{\left(\frac{\sigma_1^{\,O}}{2}\right) \pm
\sqrt{\left(\frac{\sigma_1^{\,O}}{2}\right)^2 - \sigma_2^{\,O}}} \:,
\end{equation}
which are convenient for expressing the turning points. Of course
these parameters can be obtained by making the replacement (\ref{B.36})
in Eq.~(\ref{B.10}).

We must now distinguish between the two kinds of three--dimensional
orbits occurring in the oblate cavity:

\begin{itemize}
\item {\bf $B$--orbits}

For
\begin{equation}
\label{B.40}
\sigma_2^{\,O} < 1 \qquad  {\rm and} \qquad
2\,\sqrt{\sigma_2^{\,O}} \le \sigma_1^{\,O} \le 1 + \sigma_2^{\,O} \:,
\end{equation}
the orbits always cross the focal circle. These are the orbits with a 
hyperboloidal caustic of Ref.~\cite{6} and they are analogous to  
the so--called bouncing ball modes. In this case, the
equation $p_u =0$ has four solutions. The
accessible region of phase space is 
$u_1^{\,B} \le u \le u_3^{\,B}$ and $u_4^{\,B} \le u \le u_2^{\,B}$ and 
with $0 \le v \le v_2^{\,O}$. For the lower integration limit in the $v$ 
variable
we have to take $v_1^B =0$ since the equation $p_v =0$ has no real solution in
the interval $[\,0,v_2^{\,O}\,]$. The explicit turning points  for the $u$
variable are obtained from
\begin{eqnarray}
\label{B.41}
\cos{u_2^{\,B}} & = & t_{-}^{\,O} \:, \qquad
u_1^{\,B}  =  -\,u_2^{\,B} \:, \nonumber \\
\cos{u_4^{\,B}} & = & t_{+}^{\,O} \:, \qquad
u_3^{\,B}  =  -\,u_4^{\,B} \:.
\end{eqnarray}
The constraints expressed in Eq.~(\ref{B.40}) imply the following
integration range for $\lambda_z$ and $\epsilon$,
\begin{equation}
\label{B.42}
\int d\lambda_z \quad \longrightarrow \quad 
\int_{\lambda_{-}^{\,B}}^{\lambda_{+}^{\,B}}
d \lambda_z \:,\qquad \qquad
\int d\epsilon \quad \longrightarrow \quad 
\int_{\epsilon_{-}^{\,B}}^{\epsilon_{+}^{\,B}}
d \epsilon \:,
\end{equation}
with
\begin{eqnarray}
\label{B.43}
\lambda_{\pm}^{\,B} & = & \pm\,(p_F\, \xi_{_O}) \:, \nonumber \\
\epsilon_{-}^{\,B} &=& 2\, E_F \sqrt{\sigma_2^{\,O}} \:, \qquad
\epsilon_{+}^{\,B} = E_F\,(1 + \sigma_2^{\,O}) \:.
\end{eqnarray}

\item {\bf $W$--orbits}

For
\begin{equation}
\label{B.44}
\sigma_2^{\,O} \ge 0 \qquad  {\rm and} \qquad
\sigma_1^{\,O} \ge 1 + \sigma_2^{\,O} \:,
\end{equation}
the orbits have an ellipsoidal caustic, and they are analogous to the 
so-called whispering gallery modes. For these orbits
$p^2 _u \ge 0$ for $u_1^{\,W} \le u \le u_2^{\,W}$ and $p^2_v \ge 0$ for 
$v_1^{\,W} \le v \le v_2^{\,O}$, with the corresponding turning points given by
\begin{eqnarray}
\label{B.45}
\cos{u_2^{\,W}} & = & t_{-}^{\,O} \:, \qquad
u_1^{\,W}  =  -\,u_2^{\,W} \:, \nonumber \\
\cosh{v_1^{\,W}} & = & t_{+}^{\,O} \:.
\end{eqnarray}
The integration limits for the constant of motion integrals ( cf. 
Eq.~(\ref{B.42}))
are
\begin{eqnarray}
\label{B.46}
\lambda_{\pm}^{\,W} & = & \pm\,(p_F a) \:, \nonumber \\
\epsilon_{-}^{\,W} &=& E_F\,(1 + \sigma_2^{\,O}) \:, \qquad
\epsilon_{+}^{\,W} =  E_F \left[\,\cosh{v_2^{\,O}} + 
\frac{\sigma_2^{\,O}}{\cosh{v_2^{\,O}}}\,\right]\:.
\end{eqnarray}
The integration limit $\epsilon_{+}^{\,W}$ is determined by the cavity surface.

\end{itemize}

The number of particles moving on each kind of orbits can be easily
evaluated with an integration over phase space. The total number of valence
electrons is
\beq
\label{ntot}
N=\int d \bPhi ~d{\bf I} ~F(E),
\eeq
with $F(E)$ given by Eq.~(\ref{fzero}). Clearly, setting $\hbar=1$,
\bea
N & = & 2\,\int d{\bf I}\,\theta(E_F-h_0 ({\bf I}))\:,
\nonumber \\
& = & 2\,\int_0 ^{E_F} dE \int d\lambda_z \int d \epsilon
\left|{{\partial(I_v,I_u)}\over{\partial(E,\epsilon)}}\right| ~.
\eea
We can define  $N=N_B +N_W $ with $N_B(N_W)$ the number of valence electrons 
in $B(W)$--orbits respectively and
\begin{eqnarray}
N_B &=& 2\,\int_0 ^{E_F} dE
\int_{\lambda_{-}^B(E)}^{\lambda_{+}^B (E)} d\lambda_z
\int_{\epsilon_{-}^B (E)}^{\epsilon_{+}^B (E)} d \epsilon
\left|{{\partial(I_v,I_u)}\over{\partial(E,\epsilon)}}\right|\:,
\nonumber \\
N_W &=& 2\,\int_0 ^{E_F} dE \int_{\lambda_{-}^W(E)}^{\lambda_{+}^W (E)}
 d\lambda_z
 \int_{\epsilon_{-}^W (E)}^{\epsilon_{+}^W (E)} d \epsilon
\left|{{\partial(I_v,I_u)}\over{\partial(E,\epsilon)}}\right|\:.
\end{eqnarray}
The limits for the $\lambda$ and $\epsilon$ integrals in these formulae 
are given by Eqs.~(\ref{B.43}) and (\ref{B.46}) where $E_F $ is replaced 
by $E$. Since the propagator (\ref{2.4}) has the same structure of an
integral over the classical phase space as $N$, it is convenient 
to make a similar distinction between the contribution of the two 
kinds of orbits. Thus for oblate geometry we write
\begin{equation}
D^{0}_{L'LM}(q',q,\omega)=B_{L'LM}(q',q,\omega) + W_{L'LM}(q',q,\omega)\:,
\end{equation}
with $B_{L'LM}$ and $W_{L'LM}$ still given by Eq.~(\ref{dzero}) but
with the corresponding integration limits for each kind of orbit.

The building blocks of the present calculation are the elliptic
integrals given by Eq.~(\ref{B.19}) for the prolate case. They
determine the frequencies (\ref{B.21}--\ref{B.22}), as well as the phases
(\ref{B.24}), required in the evaluation of the Fourier coefficients.
The analogous expressions for oblate geometry are obtained by
applying the transformation (\ref{B.32}) to the prolate formulae.
All formulae given for prolate geometry can be translated in the same way
for the oblate case. A little extra care should be taken for the $u$ integrals
in the case of B--orbits since  the double--well structure of the effective
potential $U_{_O} (u,\sigma_2 ^{\,O})$ makes $p_u$ become imaginary
in the interval $[u_3 ,u_4]$. This  integration range must be excluded by 
very definition of classical phase space. Then, the lower integration 
limit for $u$ in Eq.~(\ref{B.31}) becomes $u_0 =u_4^B$ for B--orbits.

%
%
%

%
%
%
%

\begin{figure}
\caption{Probability density for the natural frequencies $\{\omega_u,
\omega_v,\omega_{\varphi}\}$ at different deformations $\eta$
in the prolate sodium cluster described in text.} 
\label{NF1}
\end{figure}

\begin{figure}
\caption{$M = 0,1$ components of the photoabsorption cross section 
per valence electron for
a prolate cluster at different deformations and both in the "single--particle"
(shaded black) and "RPA--type" (shaded gray) approximations.}
\label{NF2}
\end{figure}
\begin{figure}
\caption{Fraction of particles on $B$ (full curve) and $W$ (dotted curve) orbits
as a function of deformation $\eta$ for an oblate sodium cluster.}
\label{NF3}
\end{figure}

\begin{figure}
\caption{Probability density for the natural frequencies $\{\omega_u^{W},
\omega_v^{W},\omega_{\varphi}^{W}\}$ characteristic of $W$ orbits 
at different deformations $\eta$ in the oblate sodium cluster 
described in text.} 
\label{NF4}
\end{figure}

\begin{figure}
\caption{As in Fig.~\ref{NF4} for natural frequencies characteristic of
$B$ orbits.}
\label{NF5}
\end{figure}

\begin{figure}
\caption{As in Fig.~\ref{NF2} for an oblate cluster.}
\label{NF6}
\end{figure}
\begin{figure}
\caption{Photoabsorption cross section per valence electron for  prolate
and oblate sodium clusters at several deformations. The left--hand side of the
figure shows the results obtained in a single--particle approximation
$(\sigma^{(0)})$. Arrows indicate position of plasmon peak in the spherical
case.}
\label{NF7}
\end{figure}
\end{document}